\documentclass[usenatbib]{mnras}

\usepackage{newtxtext,newtxmath}
\usepackage[T1]{fontenc}
\usepackage{ae,aecompl}

\usepackage{graphicx, subfigure}	
\usepackage{amsmath}	
\usepackage{booktabs}
\usepackage{pdflscape} 
\usepackage{longtable}
\usepackage{graphicx}
\usepackage{multirow}
\usepackage{xspace}
\usepackage{hyperref}

\newcommand{\TopoSZ}{\href{https://github.com/CMBSPEC/TopoSZ}{\tt TopoSZ}\xspace}

\graphicspath{{./figs/}}
\newcommand{\beq}{\begin{equation}}
\newcommand{\eeq}{\end{equation}}
\newcommand{\beqry}{\begin{eqnarray}}
\newcommand{\eeqry}{\end{eqnarray}}
\newcommand{\beqrys}{\begin{subequations}\begin{eqnarray}}
\newcommand{\eeqrys}{\end{eqnarray}\end{subequations}}

\newcommand{\yy}{$C_{\ell}^{yy}$\xspace}
\newcommand{\yyest}{$C_{\ell}^{\hat{y}\hat{y}}$\xspace}

\def\sec#1{{Sec.~\ref{#1}}}
\def\fig#1{{Fig.~\ref{#1}}}

\newcommand{\Planck}{{\it Planck}\xspace}
\newcommand{\id}{\,{\rm d}}
\newcommand{\qcut}{q_{\rm cut}}

\newcommand{\maskapo}{15 arcminutes}
\newcommand{\fskygal}{$f_{\rm sky}=0.50$\xspace}
\newcommand{\fskyps}{$f_{\rm sky}=0.76$\xspace}

\newcommand{\fskytotapo}{$f^{\rm G}_{\rm sky}=0.354$\xspace}
\newcommand{\fskydiffapo}{$f^{\rm uRC}_{\rm sky}=0.345$\xspace}
\newcommand{\fskyresapo}{$f^{\rm RC}_{\rm sky}=0.011$\xspace}
\newcommand{\aurc}{$A^{\rm uRC}=0.97$\xspace}

\title[A topographic $y$-map analysis]{Removing the giants and learning from the crowd: a new SZ power spectrum method and revised Compton $y$-map analysis}

\author[Rotti, Bolliet, Chluba and Remazeilles]{
Aditya Rotti$^1$\thanks{aditya.rotti@manchester.ac.uk},
Boris Bolliet$^{2,1}$\thanks{boris.bolliet@gmail.com},
Jens Chluba$^1$\thanks{jens.chluba@manchester.ac.uk},
and Mathieu Remazeilles$^1$\thanks{mathieu.remazeilles@manchester.ac.uk}
\\
$^1$ Jodrell Bank Centre for Astrophysics, Alan Turing Building, University of Manchester, Manchester M13 9PL
\\
$^2$ Columbia Astrophysics Laboratory, Columbia University, 550 West 120th Street, New York, NY, 10027, USA
}

\date{\vspace{-5mm}{Accepted 2020 --. Received 2020 --}}

\pubyear{2019}

\begin{document}

\maketitle

\begin{abstract}
The Sunyaev-Zeldovich (SZ) effect provides a powerful cosmological probe, which traditionally is approached independently as cluster number count (CNC) or power spectrum (PS) analysis.
Here, we devise a new method for analysing the $y$-map by introducing the survey completeness function, conventionally only used in the CNC analysis, in the $yy$-PS modeling. This provides a systematic method, based mainly on SZ observables, for obtaining two complementary $y$-maps, one incorporating detected/resolved clusters and the other relying only on diffuse/unresolved SZ contributions. 
We use the catalogue of clusters obtained in the \Planck CNC analysis to define the completeness function linking these two $y$-maps. The split depends on the chosen signal-to-noise detection threshold, which we vary in our discussion.
We carefully propagate the effect of completeness cuts on the non-Gaussian error contributions in the $yy$-PS analysis, highlighting the benefits of masking massive clusters.
Our analysis of the \Planck $yy$-PS for the unresolved component yields a mass bias of $b=0.15\pm0.04$, consistent with the standard value ($b\approx0.2$), in comparison to $b=0.4\pm 0.05$ for the total $yy$-PS. 
We find indications for this drift being driven by the CIB-tSZ cross correlation, which dominantly originates from clusters in the resolved component of the $y$-map. Another possible explanation is the presence of a mass-dependent bias, which has been theoretically motivated and can be quantified with our novel method.
We furthermore find first hints for the presence of the 2-halo terms in the $yy$-PS.
Finally, the proposed method provides a new framework for combining the complementary information of the CNC and PS analyses in upcoming SZ surveys.
\end{abstract}

\begin{keywords}
Galaxy clusters -- cosmology -- astrophysics
\end{keywords}

\vspace{0mm}
\section{Introduction}
\label{sec:intro}
The standard $\Lambda$CDM model of cosmology has been remarkably successful at explaining a number of observations of the Universe, made using many different cosmological probes \citep{WMAP_params, Komatsu_2011,Planck2015params}. One of the standard predictions of the model is the evolution of the number of gravitationally-bound objects as a function of cosmic time. Galaxy clusters are some of the largest gravitationally-bound structures in the Universe tracing this evolution. They are dominantly composed of dark matter, while the small fraction of baryons and their interaction with electromagnetic radiation is what allows us to directly observe them and study their properties.

The Sunyaev-Zeldovich (SZ) effect offers a unique way for studying galaxy clusters in microwave maps \citep{SZ1972}. This effect primarily entails the inverse Compton scattering of the cosmic microwave background (CMB) photons by the hot electrons inside the cluster, moving photons from low to high frequencies and introducing a specific CMB spectral distortion referred to as the $y$-distortion \citep{Zeldovich1969}.
As such, this effect was long realized to provide a powerful cosmological probe for the evolution and growth of structures in the Universe \citep[e.g.,][]{Carlstrom2002, SZreview2019}.

With relatively recent advancements in measurements of the CMB, specifically increased frequency coverage, sensitivity and angular resolution, the detection of galaxy clusters using the SZ effect has become a routine exercise \citep[e.g.,][]{Marriage2011, AdeESZCS, Bleem2015}. The SZ analysis of microwave maps is carried out using two complementary approaches. 
The {\it cluster number count} (CNC) analysis focuses on detecting individual clusters \citep[e.g.,][]{Haiman2001, Battye2003CNC}. It achieves this by using prior information on the spatial gas density profile inside clusters \citep[e.g.,][]{Arnaud2010} as well as the unique spectral signature of the clusters on the CMB spectrum. 
This is often supplemented by external information on the redshifts and masses, e.g., from optical surveys \citep[e.g.,][]{WTG2014, Hoekstra2015}. 
Studying the abundance of detected clusters at varying detection threshold can then be used to derive cosmological/astrophysical constraints \citep[e.g.,][]{Benson2011, STA11, Planck2013SZ}. 

The other complementary analysis, usually carried out at the {\it power spectrum} (PS) level, consists of constructing a Compton-$y$ parameter map, without the need to count systems. This is achieved by focusing on extracting the spatial component of the multi-frequency observations that contain the $y$-distortion spectral signature \citep{Remazeilles2011cILC}. The statistical properties of this map (e.g., 2-point or 3-point correlations) can be related to theory and thus used to derive cosmological/astrophysical constraints \citep{Komatsu_1999, Hill2013, Planck2016ymap}.
Since the map analysis itself uses only a spectral prior and does not depend on any prior information on the spatial density profiles of hot gas responsible for inducing the $y$-distortions, it can in principle yield a map of all the hot gas in the Universe \citep{Hill2015a}. It is therefore clear that the $y$-map reconstructed this way has complementary information that is not captured by the CNC modeling. In particular, the $y$-map includes both detected/resolved and diffuse/unresolved SZ contributions.

The aforementioned methods have been independently used to learn about clusters of galaxies, yielding comparable parameter constraints \citep[e.g.,][]{Bolliet:2019zuz}. One puzzling outcome is that the constraints on the amplitude of matter clustering $\sigma_8$ deduced from the \Planck SZ data is slightly lower than the one deduced from primary CMB anisotropies. This may have to do with an incomplete modelling of the ICM \citep[e.g.,][]{Henson2017, Pratt_2019}, or with the tensions between low and high redshift probes reported in the last few years \citep[e.g.,][]{Beutler2014MNRAS, Verde_2019, Douspis2019}.

Some recent works have combined the data from the PS and CNC analyses in an attempt to extract all the parameter constraining power of the SZ measurements, with some level of success \citep{hurier_lacasa_2017,Salvati2017}. These works follow the conventional wisdom of combining data sets to enhance their joint constraining power. However, it is important to realize that the CNC and PS analyses do not constrain totally independent information. The Compton-$y$ map is furthermore highly non-Gaussian, which introduces large data covariance at the PS level and thus limits the ability of SZ measurements to constrain theoretical models \citep[e.g.,][]{Komatsu_1999, Hill2013}. 
If it were possible to devise an analysis strategy that reduces this covariance at a relatively small signal cost, this could enhance the constraining power of the SZ measurements. 
Alternatively, one has to directly model higher order statistics, an approach that comes with its own challenges \citep[e.g.,][]{Bhattacharya2012, Ravenni2020}.

Indeed, one might guess that removing high peaks from the $y$-map, thereby essentially Gaussianizing the field, is a natural way forward. This counter-intuitive idea of subtracting the brightest SZ clusters to reduce errors has been proposed in previous works \citep[e.g.,][]{Komatsu_1999, Hill2013, osato2020}. These studies, however, focused on masking of clusters based on their SZ/X-ray fluxes or by introducing effective mass cuts, requiring independent measurements of the cluster field. 
Here, we revisit this idea, and propose an analysis strategy which in principle can be evaluated by only using observations of the CMB sky without the need of external measurements of the SZ field. In practice, some external ingredients still enter in relating the SZ-fluxes to the mass of the cluster, as these are usually calibrated using $X$-ray or lensing observations \citep[e.g.,][]{Arnaud2010, WTG2014} instead of fully relying on theoretical hydro-simulations that in fact make up the cluster halo-model \citep[e.g.,][]{Battaglia_2010}.

Since the refined SZ analysis strategy proposed here uses a combination of byproducts of the CNC analysis as well as the $y$-map analysis, we begin by introducing the necessary concepts involved in these two approaches (see Sect.~\ref{sec:formalism}). This allows us to naturally progress to describing the amendments to the halo-model formalism, crucial for drawing the connections between theory and the processed observations. Using this revised halo-model formalism we make quantitative theoretical evaluations that clearly highlight the SNR gains by adopting our analysis strategy. Consequently this enhances the ability of SZ measurements to differentiate between various theoretical models. With these theoretical motivations, we carry out this analysis using the SZ data products derived from \Planck measurements of the microwave sky, also highlighting some of the potential for future SZ measurements.

Current \Planck SZ analysis indicate a high mass bias, or equivalently a low $\sigma_8$ \citep{Planck2013SZ,Planck2016ymap}. Our improved SZ analysis allows us to shed light on the origin of the high mass bias observed in \Planck measurement of SZ clusters. Also, the 2-halo term \citep[e.g.,][]{Hill2013} is deemed largely unimportant for \Planck measurements of the SZ spectrum. 
Our preliminary analysis suggests that with \Planck measurements we could already be seeing a small excess of power at low multipoles, which may be due to the 2-halo contribution or possibly a diffuse SZ component \cite[e.g.][]{Hansen2005} not captured by the standard halo model. These possibilities are illuminated in Sect.~\ref{sec:results}.

\vspace{-3mm}
\section{Formalism: A topographic y-map analysis}
\label{sec:formalism}
We work within the halo model \citep[e.g.,][]{Sheth1999,Seljak:2000gq, Cooray2002}, which allows us to predict the mean number density of clusters in a given mass and redshift range, $N(M,z)$, and primarily depends on the cosmological parameters $\sigma_8 $, $\Omega_{\rm m}$ and $h$.
By adding a cluster profile to the description, one can compute both the SZ cluster number counts (CNC) and the $yy$-power spectrum (PS) observables, which then allow constraining cosmological/astrophysical parameters.

We now introduce a {\it topographic} method for analyzing the $y$-map by thinking of detected clusters as large peaks in the $y$-landscape.
This is achieved by noting the complementarity of information encoded in the CNC and PS of the $y$-map. To do this we adopt a pedagogical approach, which allows us to introduce central concepts necessary to describe this new method. We therefore begin by briefly describing the CNC analysis, which includes a discussion on the survey completeness function and the role it plays. We then discuss the main ingredients of the conventional SZ-PS analysis. Enroute we highlight essential differences in these two methods. Having developed the necessary tools, we layout the formalism for the topographic Compton-$y$ parameter analysis. Finally we highlight some of the expected benefits of this method using different theoretical estimates.

\vspace{-3mm}
\subsection{SZ cluster number counts and the survey completeness}
\label{sec:NC}
Any CMB experiment has limited sensitivity and frequency coverage, and as a consequence is only able to detect clusters above a certain cut-off in the cluster mass-redshift distribution. The CNC analysis directly relies on fitting the distribution function of the detection signal-to-noise ratio (SNR), referred to as $q$, of the Compton $Y$-parameter\footnote{$Y$ is used to denote the integrated Compton $y$-parameter of the cluster.} of the galaxy cluster as a function of its redshift: $N(q,z)$. 
Many details of the completeness modeling can be found in \citet{Planck2013SZ}.
Since the spectrum of the $y$-distortion is redshift independent, the cluster redshift is necessarily inferred from external measurements.

To connect theory to the number count observable  $N(q,z)$ requires a prescription for $q(M, z)$.
More massive clusters have a larger $Y$-parameter, and larger angular size, $\theta$, at a fixed redshift. To fix the required distributions, all observables are usually evaluated within $R_{500}$.
The prescription is thus derived by assuming a form for $\theta_{500}-M_{500}$, which is derived from a mass volume relation for a spherical cluster. It also requires a $Y_{500}-M_{500}$ relation, which is empirically derived from a sub-sample of clusters for which reliable measurements for the Compton $Y$-parameter and the mass of the cluster are available. The $Y_{500}$ is measured using the multi-matched filter (MMF) technique which requires a radial electron gas density profile for the cluster as an input \citep{Haehnelt1995,Herranz2002, Melin2006}. 
Finally, we require the MMF noise as a function of the projected angular size of the cluster $\sigma_{Y_{500}}(\theta_{500}) \equiv \sigma(\theta_{500})$, which is derived empirically from the analysis of the multi-frequency microwave observations. 
Note that this noise estimate takes into account all details of measurement noise, frequency coverage and foregrounds.

Schematically the prescription relating the halo model to observations is achieved by the following mapping:
$$N(M,z) \quad \xrightarrow[\sigma_{Y_{500}}(\theta_{500})]{\theta_{500}-M_{500},\, Y_{500}-M_{500}}\quad N(q,z).$$
It is important to note that formally, i.e., in the CNC likelihood, the counting is done in a probabilistic sense. Assuming Gaussian statistics for the MMF noise, the cumulative probability for a clusters with mass $M_{500}$ at redshift $z$ to be detected above the SNR threshold $q_{\rm cut}$ is given by \citep{Planck2013SZ},
\begin{align}
\chi(Y_{500}, \theta_{500}, q_{\rm cut},\hat{n})&=\frac{1}{2}\left[1 + \rm{erf}\left(\frac{Y_{500}/\sigma(\theta_{500},\hat{n}) - q_{\rm cut}}{\sqrt{2}}\right) \right]\,,
\nonumber
\end{align}
where the MMF filter noise, $\sigma(\theta_{500},\hat{n})$, changes as a function of sky location $\hat{n}$, owing to changing foregrounds and any non-uniform observing depth of the observations. While it is in principle possible to use the full error information, the cosmological analysis can be simplified by defining a sky-averaged completeness function, which can be obtained by integrating over all sky patches on which the MMF noise is evaluated:
\begin{align}
\bar{\chi}(M,z, q_{\rm cut}) &= \frac{\int \chi(Y_{500}, \theta_{500}, q_{\rm cut},\hat{n}) \id \Omega_{\rm masked}}{  \int \id \Omega_{\rm masked}}.
\end{align}
Note that the resultant survey completeness function takes into account the error on the inferred detection SNR of a cluster. In addition, the \Planck number count analysis also assumes the Compton-$Y$ parameter for clusters to have an intrinsic scatter. For this, the $Y$ parameter is assumed to follow a log-normal field and the Gaussian width of $\ln Y$ is estimated while empirically fitting the $Y_{500}-M_{500}$ relation. These factors make the detection SNR assigned to a cluster fuzzy and are taken into account in the \Planck CNC likelihood function. 

While other details of the \Planck number count analysis are important and interesting in their on right, for our discussion below, the average survey completeness function is the crucial ingredient we will borrow to formulate a topographic $y$-map power spectrum analysis. It will essentially enter as a weight factor in the halo mass function which amends the evaluation of the $yy-$power spectrum, as we elucidate in the following section.

\subsection{Compton $yy$ power spectrum analysis}

A map of the Compton $y$-parameter is composed of the cumulative signal from all clusters and diffuse hot gas in the Universe. It can be extracted from multi-frequency microwave observations using an ILC algorithm \citep[e.g.,][]{Planck2013ymap} and can similarly be used to draw inferences on cosmological and astrophysical parameters \citep[e.g.,][]{Bolliet2017, Salvati2017}. 
The usual power spectrum analysis is based only on the contributions from collapsed halos, which we focus on here, considering both the one and two halo terms \citep[see][for additional details]{Molnar2000, Komatsu2002b, Hill2013}.
Additional contributions from filaments and bridges are not accounted for but could play a role at the largest scales \citep[e.g.,][]{Hansen2005} as we also discuss below.

As for the SZ number counts, the halo model is used to compute the theoretical Compton $y$-map power spectrum. The 1-halo term can be estimated using the following expression,
\begin{align}
C_{\ell}^{yy,{\rm 1h}} 
&=\int_{0}^{z_{\rm max}} {\rm d}z \frac{{\rm d}V}{{\rm d}z}\int_{M_{\rm min}}^{M_{\rm max}} {\rm d}M \frac{{\rm d}N}{{\rm d}M{\rm d}V} \,|y_{\ell}(M,z) | ^2
\nonumber \\
&\equiv \langle |y_{\ell}(M,z) | ^2 \rangle,
\label{eq:yy_total}
\end{align}
where $\id N/(\id M\id V)$ is the halo mass function, which determines the comoving number density of halos of a given mass $M$ at each redshift $z$, and $y_{\ell}$ is the 2d Fourier transform of the pressure profile projected along the line-of-sight. We also introduced the short-hand notation, $\langle X \rangle$, for the average of the quantity $X$ weighted by the distribution of the number of halos as a function of mass and redshift\footnote{Note that $\id \id N(M,z)/(\id M\id V)$ can be thought of as a probability distribution function, however it is not normalized to unity}.
The resultant PS thus includes contribution from clusters at all redshifts and for all masses. Details regarding the implementation of $\id N(M,z)/(\id M\id V)$ and $y_{\ell}(M,z)$ can be found in \citet{Bolliet2017}. Notably, following the \Planck SZ analysis, we use \citet{Tinker2008} for the halo mass function with mass bias $1-b=0.8$ [i.e., $B=1/(1-b)=1.25$] and the pressure profile from \citet{Arnaud2010}, which was also applied in the cluster finding algorithm of the original \Planck CNC analysis.

In addition to the 1-halo term described above, the 2-halo contributions arising from correlation between the spatial position of clusters are present and become relevant on large angular scales ($\ell \lesssim 100$). 
%
%
For simplicity we use the Limber approximation to estimate this contribution. The 2-halo contribution can then be expressed as \citep[see Appendix of][]{Hill2013}\,,
\begin{subequations}
\begin{align}
C^{yy,{\rm 2h}}_{\ell}
&\approx \int_{0}^{z_{\rm max}} {\rm d}z \frac{{\rm d}V}{{\rm d}z}\,\langle b_{\rm h} \,|y_{\ell}| \rangle_M^2\,P_{\rm lin}\left(\frac{\ell+1/2}{d(z)}; z\right) \,,
\\
\langle b_h \,|y_{\ell}|\rangle_M &=\int_{M_{\rm min}}^{M_{\rm max}} {\rm d}M \frac{{\rm d}N }{{\rm d}M{\rm d}V} \,b_{\rm h}(M,z) \,|y_{\ell}(M,z)|,
\label{eq:yy_total}
\end{align}
\end{subequations}
where $P_{\rm lin}(k,z)$ denotes the linear matter power spectrum, $b_{\rm h}(M,z)$ is the halo bias \citep[][]{1984ApJ...284L...9K,Bardeen:1985tr, Mo_1996,2001MNRAS.323....1S,Dalal_2008,Tinker2010} and  $d(z)$ is the comoving distance. We refer to \cite{Komatsu_1999} and \cite{Hill2013} for the use of the halo bias in the context of the $yy$ power spectrum.

We compute the linear matter power spectrum, and cosmological distances, using {\tt CLASS} \citep{lesgourgues2011cosmic,Blas2011} within \verb|CLASS_SZ| \citep{Bolliet2017}. For the halo bias we use the \citet{Tinker2010} formula (see their Eq. 6 and Table 2), which gives the bias in terms of the `peak height' $\nu=\delta_c/\sigma(M)$, where $\delta_c$ is the critical overdensity for collapse and $\sigma(M)$ the variance of the matter overdensity field  smoothed over a sphere whose size corresponds to the typical cluster size, $R=(3M/4\pi\rho_\mathrm{m})^{1/3}$. In this model, the bias is unity at low $\nu$ (i.e., when clustering is \textit{efficient}, typically for less massive halos) and increases quickly for larger $\nu$ (i.e., larger masses). %
As we discuss in more detail below, the 2-halo term usually only contributes at the level of ten percent to the total \yy at $\ell \lesssim 100$, but more interesting is that its relative contribution increases significantly on progressively removing the brightest clusters detected in the SZ survey.

\begin{figure}
\centering
\includegraphics[width=0.95\columnwidth]{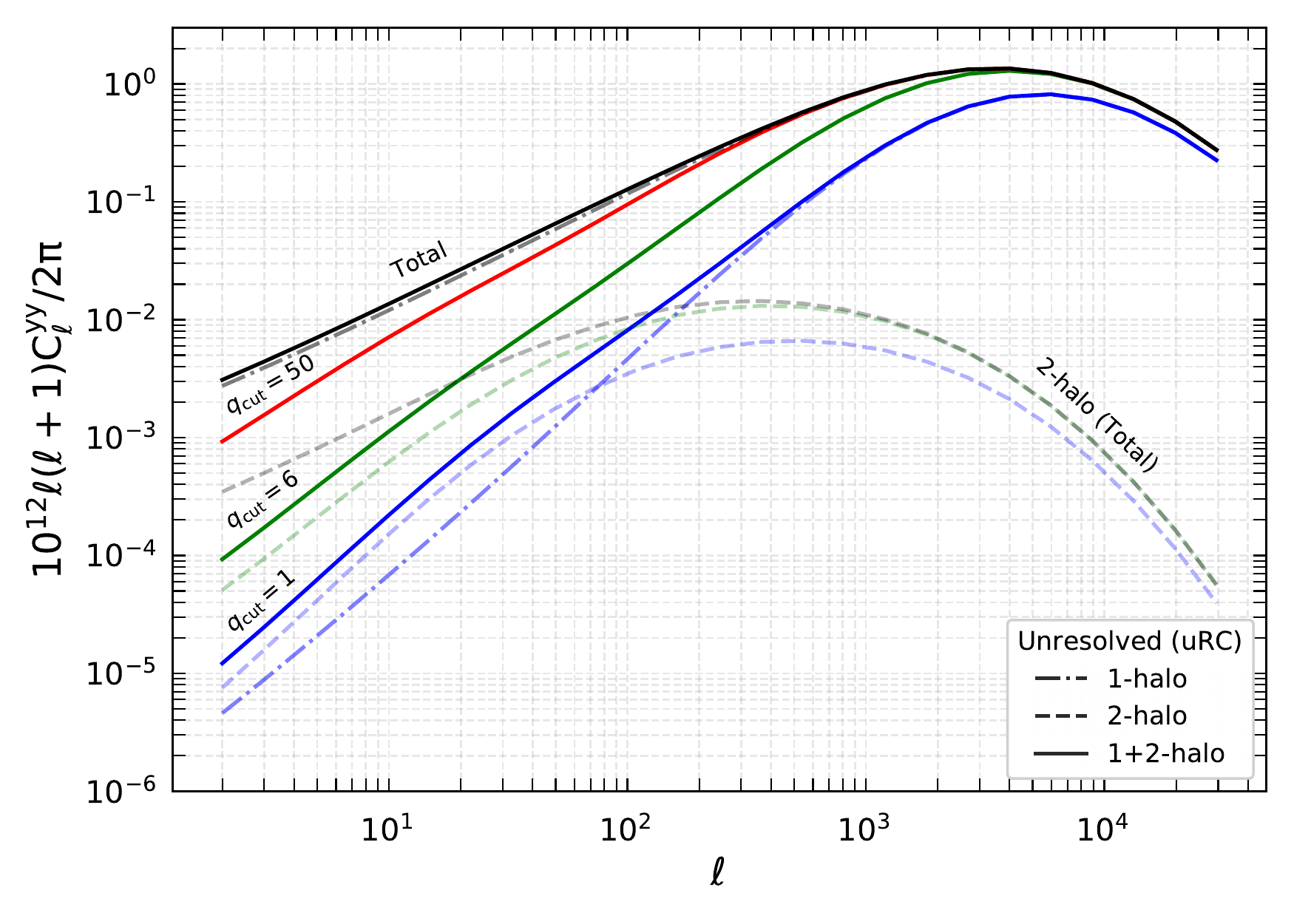}
\\[-1mm]
\includegraphics[width=0.95\columnwidth]{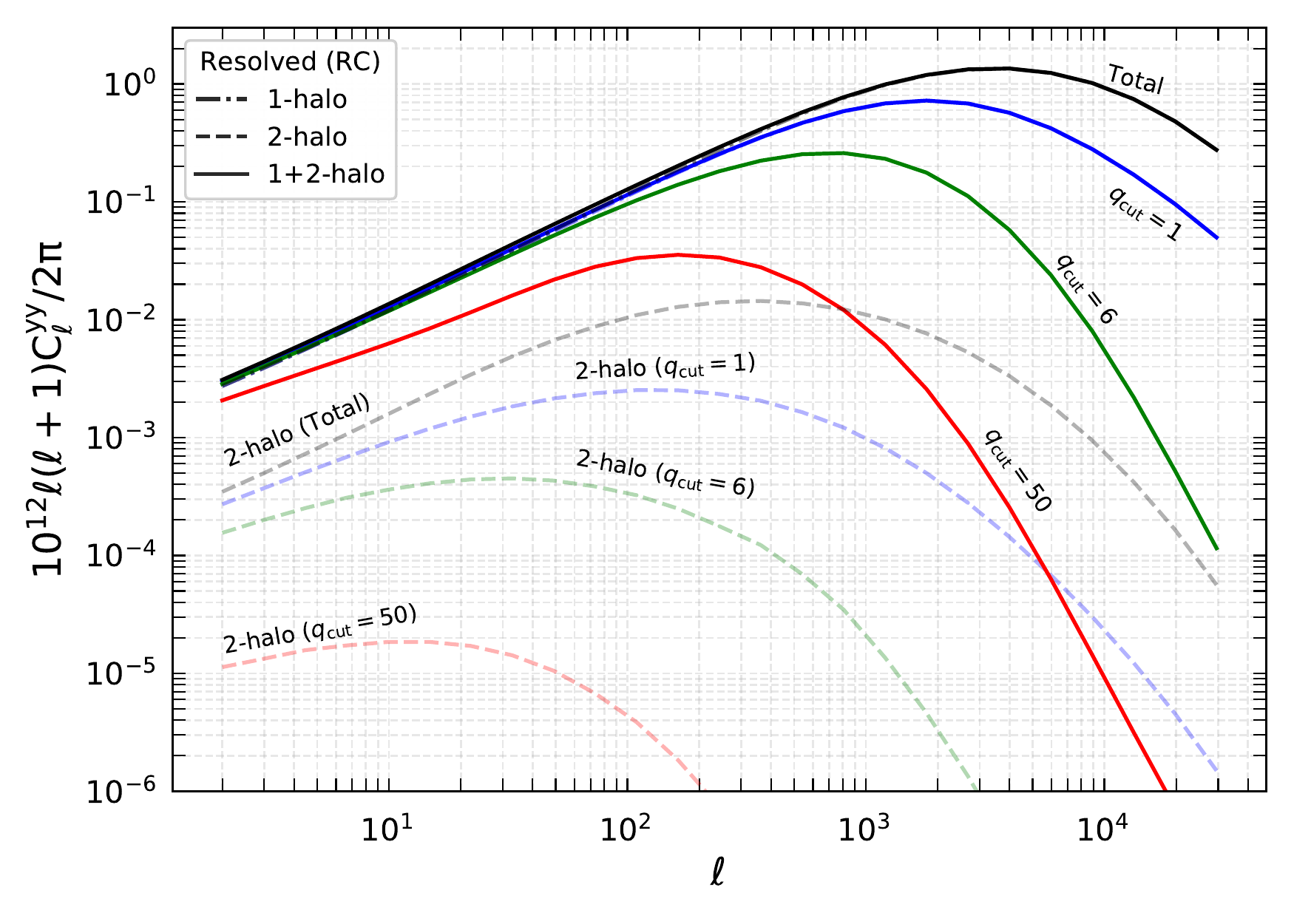}
\vspace{-3mm}
\caption{The theoretical $yy$-power spectrum augmented by explicit completeness modeling.
The dot-dashed lines depict the 1-halo contribution, while the dashed lines indicate the 2-halo contribution to \yy. The solid lines denote the total \yy which includes both the 1-halo and 2-halo contributions. The top panel illustrates this decomposition for uRC of the $y$-map and the bottom panel shows the same for RC of the $y$-map. Note that for low $q_{\rm cut}$, the 2-halo term has a dominant contribution for the uRC while for RC the 2-halo contribution is always sub-dominant.}
\vspace{-3mm}
\label{fig:1halo2halo}
\end{figure}

\vspace{-3mm}
\subsubsection{Power spectrum completeness modeling}\label{ss:psmod}
The conventional PS analysis does not explicitly account for the limited experimental sensitivity in its theoretical modeling as it is done in the CNC analysis. This is a valid approach since it is fair to assume that noise in the \yy measured from the data naturally accounts for this limitation. As such, it is one of the manifest differences between source counting and analyses that rely on characterizing the $n$-point statistics of a field.
However, through masking one can remove clusters from the maps and thus subtract their PS contribution. In this step, mass-cuts can usually not be implemented easily and furthermore necessitate independent redshift information.
Thus, a more practical way to account for masking effects in the theoretical PS modeling is to perform the separation using a specified $\qcut$ together with a CNC completeness model, as we now elucidate.

In the brief recap of the CNC analysis in Sect.~\ref{sec:NC}, we introduced the sky-averaged survey completeness function $\bar\chi$, which we now combine with the PS analysis to perform a $\qcut$-dependent $yy$-power spectrum modeling. Note that this invariably relies on having evaluated parts of the number count analysis to yield a cluster catalogue with the detection SNRs for each cluster,  a $Y_{500}-M_{500}$ scaling relation and the survey completeness function. 
Given this data, the $y$-map can now be divided into two parts, one composed of clusters above the preset SNR threshold $\qcut$ and one with $q$ below. We will refer to these as resolved component (RC) and unresolved component (uRC) of the $y$-map respectively. 
While this reference is only accurate for a reasonably low $\qcut$, for simplicity we persist with it for all $q_{\rm cut}$ considered in this work, owing to the lack of a better nomenclature\footnote{Note that for sufficiently low $q_{\rm cut}$, one can more simply think of the RC as detected/resolved SZ sources, which can be directly modeled using number count methods.}.

To compare with theory, the RC and uRC $yy$-power spectra are given by the following expressions:
\begin{subequations}
\begin{align}
C_{\ell}^{yy,{\rm 1h}}(\qcut) &=  \big\langle |y_{\ell}(M,z) | ^2 \Phi\left(M,z , q_{\rm cut}\right) \big\rangle,
\\
\Phi\left(M,z,q_{\rm cut}\right)&=\begin{cases}
\bar{\chi}\left(M,z, q_{\rm cut}\right) & \mathrm{RC}\\
1-\bar{\chi}\left(M,z, q_{\rm cut}\right) & \mathrm{uRC}
\end{cases}
\end{align}
\end{subequations}
Hence, according to this model when $q_{\rm cut} \rightarrow \infty$ and $\bar{\chi} \rightarrow 0$, the uRC power spectrum becomes the total PS and the RC power spectrum vanishes: this corresponds to the case where {\it none} of the halos are masked. On the contrary, when $q_{\rm cut} \rightarrow 0$ and $\bar{\chi} \rightarrow 1$, the RC power spectrum becomes the total and the uRC vanishes: this corresponds to the case where {\it all} of the halos are masked\footnote{Note that in practice it is not possible to take this limit on the data since as for very low $q_{\rm cut}$ it becomes impractical to construct a reliable mask.}. 

For the 2-halo contribution, we include the completeness in a similar way: 
\begin{align}
C^{yy,{\rm 2h}}_{\ell}
&\approx \int_{0}^{z_{\rm max}} {\rm d}z \frac{{\rm d}V}{{\rm d}z}\,\langle b_h \,|y_{\ell}|\,\Phi\rangle_M^2\,P_{\rm lin}\left(\frac{\ell+1/2}{d(z)}; z\right),
\label{eq:2h_yy_total_mod}
\end{align}
which naturally follows from thinking of the completeness cuts as part of the halo-model, $\id N^*/(\id M \id V)=\Phi\id N/(\id M \id V)$.

To illustrate several key points, as an example, we use the \Planck completeness function to evaluate the modified halo model. For these illustrations we assume a spatially-flat $\Lambda$CDM cosmology with $\sigma_8 = 0.8$, $\Omega_\mathrm{b} = 0.05$, $\Omega_\mathrm{cdm} = 0.27$, $h = 0.7$ and three degenerate massive neutrinos with $\Sigma m_\nu=0.06$ eV. It is important to bear in mind that the key features of the results, which we now discuss, depend only on the fact that the high $q$ clusters are removed from uRC and therefore should not have any critical dependence on the specifics of the survey completeness function.
The resultant PS is illustrated in Fig.~\ref{fig:1halo2halo} for various values of $\qcut$. For completeness, we show both the 1- and 2-halo contributions. Varying the $\qcut$, we observe that the uRC power spectrum is affected the most on large angular scales, while the RC power spectrum changes mostly at small angular scales. This immediately highlights that completeness modeling affects the contributions from massive clusters for the uRC and low mass halos in the RC, a point that we will address more rigorously in Sect.~\ref{sec:mean_masses}.

Looking at Fig.~\ref{fig:1halo2halo}, it is important to notice that for the total $yy$ power spectrum, the 2-halo term only contributes at the level of a few percent to the total $yy$-power spectrum at $\ell \lesssim 100$. However, varying the value of $\qcut$, we observe that the relative contribution significantly changes for the uRC power, with the 2-halo contribution becoming nearly equal to the 1-halo at $q_{\rm cut}=6$ and even being the dominant contribution for lower values of $q_{\rm cut}$. The effect is most significant at large angular scales, $\ell \lesssim 100-200$. For the RC power spectrum, the 2-halo term always remains subdominant. For a $\qcut$-dependent analysis it is thus more important to carefully include the contributions from the 2-halo term for the uRC, and we will return to discussing this point again in Sect.~\ref{sec:2_halo}.

It is also important to realise that all these features are likely to have some dependence on details of the completeness function used to evaluate the halo model. In addition, the relevance of the 2-halo term needs to be compared to contributions from diffuse SZ effect, e.g., due to filaments and bridges, which also contribute at similar level on the largest angular scales \citep[e.g.,][]{Hansen2005}.

\begin{figure}
\centering
\includegraphics[width=0.98\columnwidth]{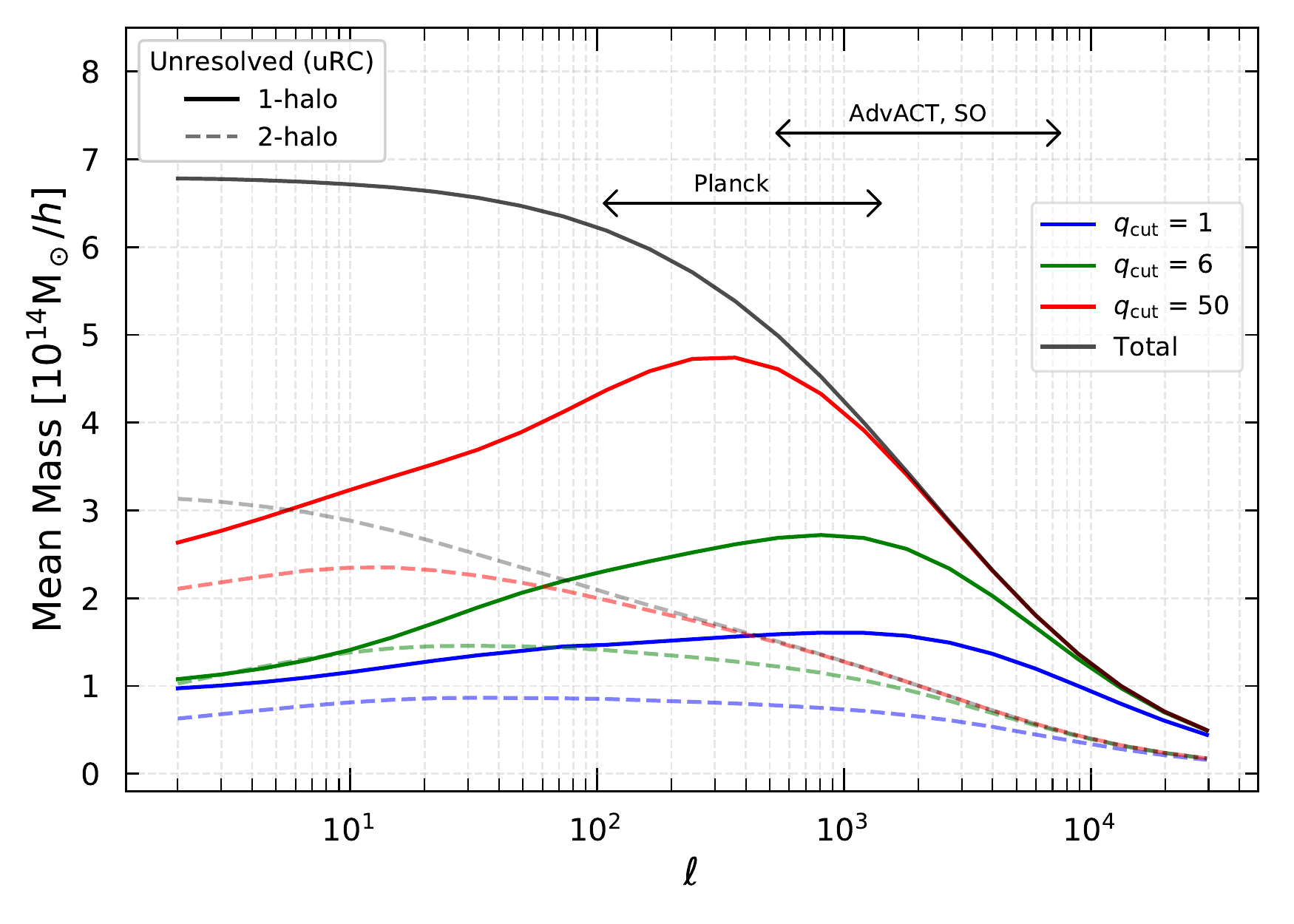}
\\[-0.5mm]
\includegraphics[width=0.98\columnwidth]{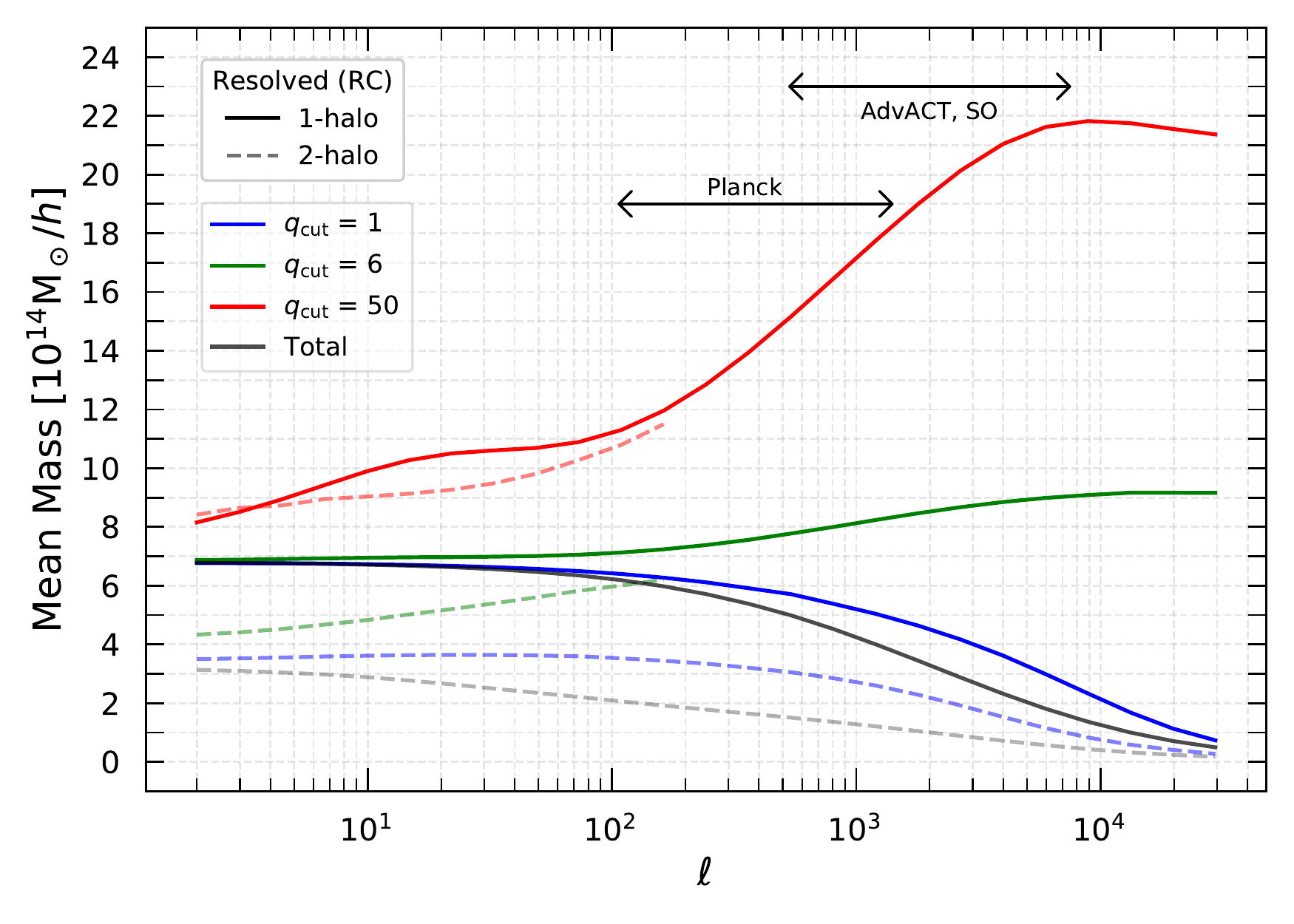}
\vspace{-2mm}
\caption{This figure depicts the mean mass of clusters contributing to each multipole in the $yy$-power spectrum. For the uRC (upper panel), high mass systems dominantly contribute at low multipoles while less massive clusters dominantly contribute at higher multipoles. For the RC (lower panel), the picture is complicated by normalization effects (see text for discussion).}
\vspace{-2mm}
\label{fig:mean_mass_contrib_multipole}
\end{figure}

\vspace{-0mm}
\subsubsection{Mean masses for various values of $\qcut$.}
\label{sec:mean_masses}
The \Planck completeness function is defined in terms of SNR thresholds and not in terms of the cluster mass. To understand the $q-M$ correspondence, we thus define the mean mass for both 1-halo and 2-halo contributions to the SZ power spectrum after integrating over redshift and cluster mass (see Fig.~\ref{fig:1halo2halo}). For the 1-halo term, the mean mass can be estimated using
\begin{align}
\langle M \rangle_{\ell}
&= \frac{\langle M |y_{\ell}(M,z) | ^2 \Phi\left(M,z , q_{\rm cut}\right)\rangle}{\langle |y_{\ell}(M,z) | ^2 \Phi\left(M,z , q_{\rm cut}\right) \rangle},
\label{eq:M_total_ell}
\end{align}
where we simply think of the differential power spectrum contributions, $\id \ln C^{yy}_{\ell}/\id M$ as a probability distribution function, as in Eq.~(25) of  \cite{Komatsu2002b} (see their Figure 6).

Similarly for the 2-halo contribution to \yy, the simplest way to estimate the dominant mass contribution is to use the replacement $\langle b_h \,|y_{\ell}| \,\Phi \rangle_M \rightarrow \langle M b_h \,|y_{\ell}| \,\Phi \rangle_M$ in Eq.~\eqref{eq:2h_yy_total_mod}. With this one can compute the mass correlation function
\begin{align}
\langle M M' \rangle_{\ell}
&= 
\frac{\int_{0}^{z_{\rm max}} {\rm d}z \frac{{\rm d}V}{{\rm d}z}\,\langle M b_h \,|y_{\ell}|\,\Phi\rangle_M^2\,P_{\rm lin}\left(\frac{\ell+1/2}{d(z)}; z\right)
}{\int_{0}^{z_{\rm max}} {\rm d}z \frac{{\rm d}V}{{\rm d}z}\,\langle b_h \,|y_{\ell}|\,\Phi\rangle_M^2\,P_{\rm lin}\left(\frac{\ell+1/2}{d(z)}; z\right)
}.
\label{eq:MM'_total_ell}
\end{align}
We emphasize that in both Eq.~\eqref{eq:M_total_ell} and \eqref{eq:MM'_total_ell} the completeness function is part of the normalization condition.

In Fig.~\ref{fig:mean_mass_contrib_multipole}, we show $\langle M \rangle_{\ell}$ and $\sqrt{\langle M M' \rangle_{\ell}}$ for the uRC and RC. Focusing on the uRC (upper panel of Fig.~\ref{fig:mean_mass_contrib_multipole}), we see that for the total $yy$ power spectrum massive clusters mostly contribute at large angular scales, while the small scales are dominated by lower mass systems. The mean mass of clusters contributing to the low multipoles reduces significantly on lowering the value of $\qcut$. Similarly, at intermediate angular scales most relevant to the \Planck analysis ($\ell \simeq 10^2-10^3$), one observes a significant drop in the mean mass with $\qcut$. These statements are true for both the 1- and 2-halo contributions and are naturally expected as the most massive systems are progressively masked. Advanced ACT \citep{Henderson2016advACT} and SO \citep{SOWP2018} will increase the relevant multipole-range to $\ell \simeq 10^4$, probing even smaller masses.

Turning to the RC (Fig.~\ref{fig:mean_mass_contrib_multipole}, lower panel), we see that in particular the effective mass at small angular scales increases dramatically with the chosen value for $\qcut$. This indicates that only the most massive but distant (i.e., small scale) cluster systems contribute.
With an increasing $\qcut$ only the most massive systems remain in the RC, and it becomes progressively inappropriate to perform a power spectrum analysis on this component of the $y$-map\footnote{In principle, however, there is no issue, since the 1-halo contribution to the power spectrum is literally composed of combining single clusters ringing in harmonic space on the sky.}. At large angular scales, the variation of the effective mass with $\qcut$ remains more moderate, highlighting that these contribution come from close by. 
We note that normalization effects are more pronounced for the RC than for the uRC. This stems for the fact that due to the completeness modeling the denominators of Eq.~\eqref{eq:M_total_ell} and \eqref{eq:yytri_total} can become very small for the RC with increasing $\qcut$.

\vspace{-2mm}
\subsection{Theoretical modeling uncertainties}
\label{sec:tri_est}
\begin{figure}
\centering
\includegraphics[width=\columnwidth]{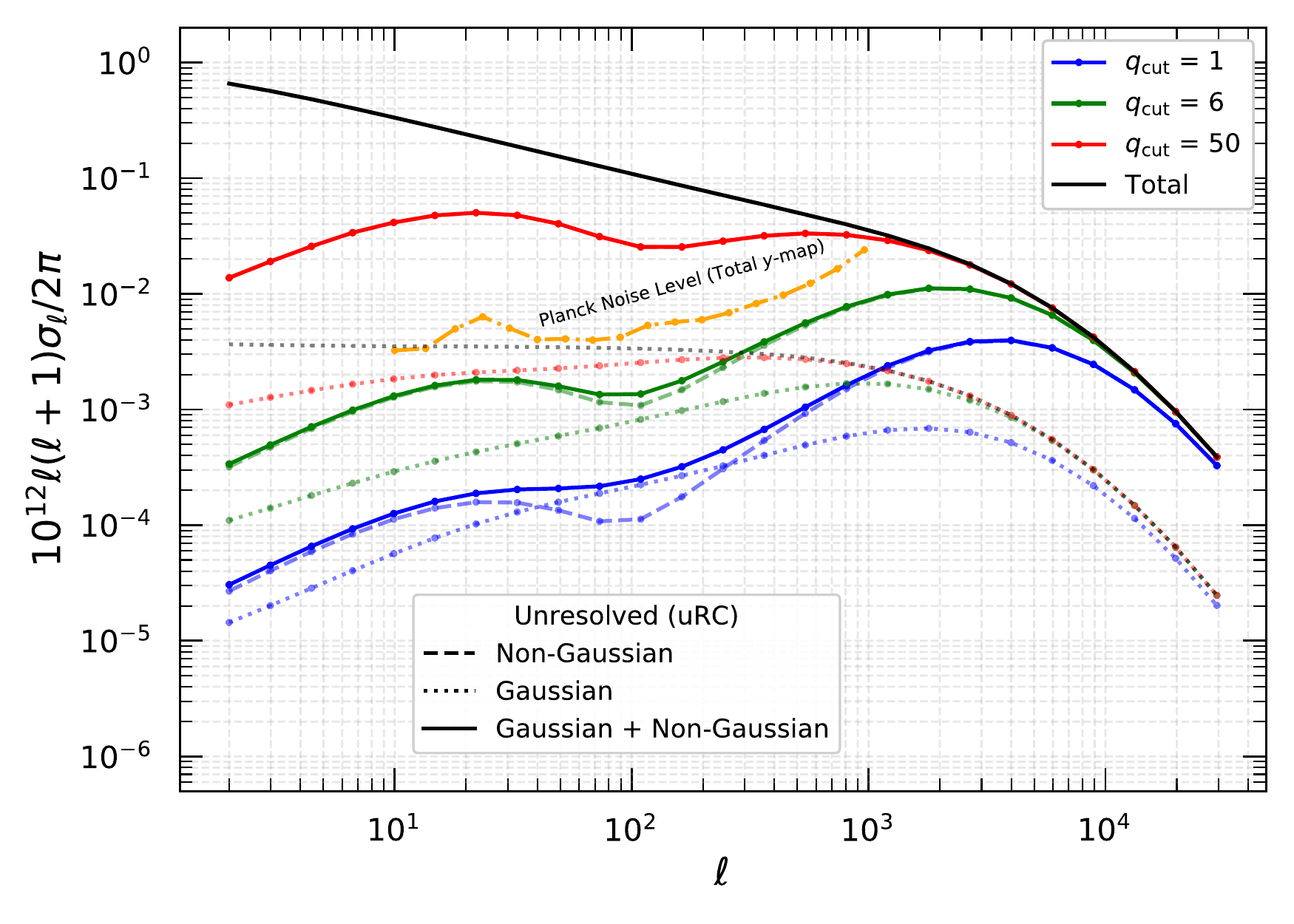}
\\[2mm]
\includegraphics[width=\columnwidth]{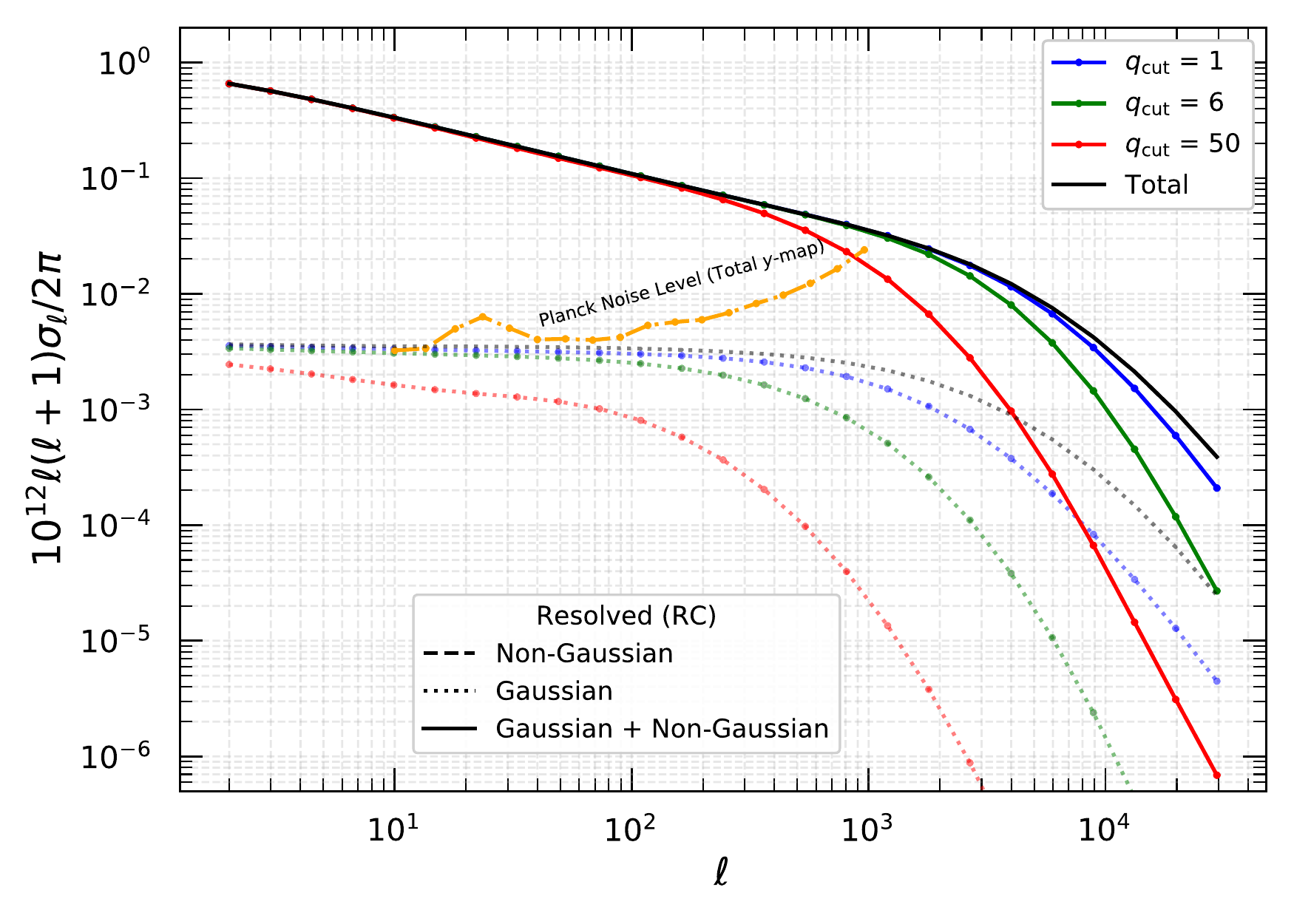}
\vspace{-2mm}
\caption{Theoretical uncertainties in the $yy$-PS computation. 
The dashed lines depict the diagonal of the trispectrum error covariance, the dot-dashed lines depict the covariance of the Gaussian part and the solid lines show the total. While the top panel shows the error decomposition for uRC, the bottom panel illustrates the same for the RC of the $y$-map. 
Note that for this figure we used $f_\mathrm{sky}=0.354$, and that the errors are binned: a dot marker indicates the centre of each bin. 
The error estimated from \Planck total $y$-map is shown for reference.
}
\vspace{-0mm}
\label{fig:yy_err_qcut}
\end{figure}

For the SZ power spectrum analysis, the measurement noise and Gaussian cosmic variance form only a part of the noise budget, and these are usually estimated directly from observations. 
In the conventional \yy analysis, these form a sub-dominant contribution to the total noise budget at multipoles below a few hundred, since the $y$-map is highly non-Gaussian. The non-Gaussian component of the noise is estimated by evaluating the trispectrum $T^{yy}_{\ell \ell'}$ \citep[e.g.,][]{Cooray2001, Komatsu2002b, Hill2013}, 
\begin{gather}
T^{yy}_{\ell \ell'}
=\frac{1}{4 \pi} \int \!\!\id z \frac{\id V}{\id z}\! \int \!\!\id M \frac{\id N(M,z)}{\id M\id V} |\tilde{y}_{\ell}(M,z)|^2 |\tilde{y}_{\ell'}(M,z)|^2\,,\nonumber 
\\ 
\equiv \frac{\langle |\tilde{y}_{\ell}(M,z)|^2 |\tilde{y}_{\ell'}(M,z)|^2 \rangle}{4 \pi},
\label{eq:yytri_total}
\end{gather}
where we again use the Limber approximation \citep[e.g.,][for more discussion]{Hill2013}.
In the cosmic variance limited case, the corresponding non-Gaussian noise contribution dominates the total theoretical noise budget at all multipoles, as seen in Fig.~\ref{fig:yy_err_qcut}. As alluded to before, this significantly degrades the constraining power of the \yy analysis but can be amended by reducing the non-Gaussian error contribution \citep[e.g.,][]{Hill2013}. When augmenting the PS analysis with the survey completeness function, it is important to also update the trispectrum error by similarly incorporating the completeness function in its evaluation. We can therefore define the appropriate trispectrum error covariance as
\begin{align}
T^{yy}_{\ell \ell'}(\qcut)&= \frac{\big\langle |\tilde{y}_{\ell}(M,z)|^2 |\tilde{y}_{\ell'}(M,z)|^2  \Phi\left(M,z , q_{\rm cut}\right) \big\rangle}{4 \pi},
\end{align}
which for $\qcut \rightarrow \infty$ reduces to the usual trispectrum contribution for the uRC (while for the RC it vanishes).

In Fig.~\ref{fig:yy_err_qcut} we illustrate the corresponding diagonal of the non-Gaussian error, $\propto (T^{yy}_\ell/f_\mathrm{sky})^{1/2}$, in comparison to the Gaussian cosmic variance contribution. Here, $f_{\rm sky}$ denotes the sky fraction covered by the $y$-map after masking. For illustration, we binned the errors in logarithmic bins with width $\Delta\ln \ell=0.4$. Thus, the number of modes in each bin is $N\simeq 2\ell\Delta \ell\simeq \ell^2 \Delta \ln \ell \propto \ell^{2}$. Since $C^{yy}_\ell\simeq 1/\ell$ (cf. \fig{fig:1halo2halo}), the binned Gaussian cosmic error, specifically $\ell(\ell+1) \sigma_{\ell} \simeq \ell^2 C_{\ell} ^{yy}/ N^{1/2}$, is nearly constant at large scales (see \fig{fig:yy_err_qcut}). The errors corresponding to low value of $q_{\rm cut}$ show a suppression of power at low multipoles, as removal of high mass clusters preferentially reduces the power on large scales. 

For the uRC, the non-Gaussian contribution drops significantly with the $\qcut$. For $\qcut\lesssim 3$, we even find that the Gaussian contributions start dominating in a small multipole window around $\ell \simeq 10^2$. This clearly suggests that the analysis of the uRC can theoretically improve the constraining power of the $yy$-PS for cosmology (we will quantify this in Sect.~\ref{sec:total_error_est}). However, even if the plain instrumental PS noise is subdominant,  foreground residuals and marginalization play a crucial role in the discussion (see Fig.~\ref{fig:qcut_err_spec} for $\qcut=6$).
For the RC, on the other hand, the non-Gaussian error contributions always remain dominant, strongly hindering cosmological inference.

It is important to mention that we omitted non-Gaussian variance contributions from clustering  \citep[i.e., the so-called super sample covariance, see][]{osato2020}. Their evaluation is beyond the scope of this paper, which is a first attempt of propagating cluster masking to the PS analysis using actual data. We expect this to mainly affect the error bars, but not the main aspects of the conclusions\footnote{While we were preparing this manuscript for submission, \cite{osato2020} presented a thorough theoretical treatment of the super sample covariance for the tSZ power spectrum, including the effect of masking. Their finding shows that the error bar on the mass-bias  measured by a  masked $y$-map power spectrum analysis can be affected at the level of several percents. This is a small but potentially non-negligible effect which will be addressed in future work.}.

\vspace{-3mm}
\subsection{Interplay with cluster number counts}
\label{sec:PSvsCNC}
We close our theoretical considerations by again highlighting some of the interplay of the proposed PS modeling with the usual CNC method.
Firstly, the RC essentially accounts for a sub-sample of clusters that are considered in the CNC analysis, but in this case by using map-based methods. However, given the highly non-Gaussian nature of the RC $y$-field, for large $\qcut$ it is more sensible to use CNC methods to extract information. 

    Secondly, to perform our $\qcut$ power spectrum analysis we first need to build the statistics for the CNC analysis, as already noted above. As such, the CNC analysis can naturally be used to deal with high SNR systems, while the uRC is best-modeled using power spectrum methods. The covariance between these two observables is indeed expected to be small, which provides a simple and powerful avenue for combining both likelihoods for existing and future data. However, as we showed in Fig.~\ref{fig:mean_mass_contrib_multipole}, the systems relevant to the RC and uRC are quite different, such that also astrophysical properties (e.g., mass-bias) generally have to be dealt with independently.

Finally, the uRC also naturally includes contributions from diffuse $y$. The halo model only accounts for collapsed systems and thus cannot capture information from filaments and bridges. Their effect can become particularly relevant at large angular scales, with a potential to bias cosmological inferences, as the \yy modeling omits them.

\vspace{-3mm}
\section{From microwave observations of the CMB sky to a topographic $y$-map analysis}
\label{sec:spectral_analysis}
In the previous section we illustrated, using theoretical estimates, that removing some of the brightest clusters from the $y$-map should enhance the ability of the $yy$-power spectrum to constrain model parameters. In this section, we begin by laying out all the algorithmic steps involved in carrying out the proposed topographic analysis, starting right from the multi-frequency measurements of microwave maps. We end this section by describing details of the map level and power spectrum analysis using actual \Planck data. The parameter constraints derived from this topographic analysis of the \Planck SZ data are discussed in Sec.~\ref{sec:mla}.

\vspace{-3mm}
\subsection{A general algorithm} \label{sec:topo_algo}
Performing the proposed topographic analysis starting from the multi-frequency observations of the CMB sky essentially consists of the following algorithmic steps: 
\begin{itemize}

\item Reconstructing the $y$-map from the multi-frequency measurements using a component separation algorithm (e.g ILC). 

\item Running the cluster detection algorithm (e.g. MMF) on the multi-frequency maps and tabulating the SNR with which each cluster is detected. Here it is necessary to estimate the characteristic size of the cluster, $\theta_c$, at which the detection SNR is maximized, as this will be used for masking the cluster. It is also important to estimate the MMF noise as a function of the filter size on different portions of the sky as this will be used to define the survey completeness function. 

\item Using the obtained cluster catalogue to construct a mask, that excludes clusters detected above some preset detection threshold $q_{\rm cut}$. Here, $\theta_c$ determines the size of the mask surrounding each cluster. This mask will be used to slice the $y$-map into RC and uRC 

\item Evaluating the \yyest corresponding to each slice of the $y$-map, duly corrected for the partial sky coverage, owing to foreground contamination and measurement footprint.
Here one also estimates the Gaussian component of the variance on \yyest, which includes the instrument noise variance.

\item Revising the Compton-$y$ power spectrum likelihood to include the completeness function used in the number count analysis. This essentially involves an update to the evaluation of \yy and the trispectrum covariance $T_{\ell \ell'}^{yy}$, which accounts for the completeness function $\Phi$, as detailed in Sec.~\ref{sec:formalism}.

\item Finally, deriving cosmological constraints from the uRC as well as the RC of the $y$-map, where this procedure can be performed, for different values of $q_{\rm cut}$. 

\end{itemize}
The topographic $y$-map analysis proposed here in principle does not rely on using external measurements and can be completely evaluated using only multi-frequency microwave observations as input.
However, as mentioned above, the theoretical PS modeling required for the analysis still contains data-driven ingredients related to scaling relations and cluster profiles.
Note that once these data products are derived, it is easy to repeat the topographic analysis for different values of $q_{\rm cut}$.
These seemingly redundant analyses act as consistency checks and increase the potential of unravelling the need for a revised theory or systematics in the data. For instance, in an ideal setting (i.e. when we have data which perfectly matches the theory model predictions), the expectation is that a model parameter analysis on different decompositions of the $y$-map should yield compatible constraints (with a certain $\qcut$ optimizing the error bars). Any deviations from this simple expectation would be indicative of either systematics in the data that are poorly understood or potentially an incomplete theory (e.g, mass dependent hydrostatic mass bias) or even a combination of the two. So while we arrived at this analysis method motivated by its improved constraining power, this framework also provides these additional diagnostic benefits.

\subsection{A topographic analysis of the \Planck $y$-map}
\Planck measurements of the microwave sky are already sensitive to the measurement of the $y$-distortions and have therefore successfully delivered maps of the Compton $y$ parameter \citep{Planck2016ymap}. 
The CNC analysis was carried out by the \Planck collaboration and the resultant catalogue of clusters detected using a few different analysis pipelines are also available\footnote{All \Planck compact object catalogues can be accessed at this link: \href{https://irsa.ipac.caltech.edu/data/Planck/release_2/catalogs/}{https://irsa.ipac.caltech.edu/data/Planck/release\textunderscore2/catalogs/}}. 

Following the \Planck number count analysis \citep{Planck2013SZ}, here we work with the MMF3 cluster catalogue \citep{planck2016_cnc_cat}. From this catalogue we specifically use the estimates of cluster detection SNR, the estimate of the SZ mass and the cluster sky coordinates, with details to follow. The other important byproduct of the MMF analysis is the filter noise as a function of the characteristic size of the cluster $\theta_c$, which is available as a part of the \href{https://github.com/cmbant/CosmoMC}{\tt CosmoMC} package\footnote{The MMF noise details are encoded in the files: SZ\textunderscore skyfracs.txt, SZ\textunderscore ylims.txt \& SZ\textunderscore thetas.txt, which can be found inside the "data" folder of the \href{https://github.com/cmbant/CosmoMC}{{\tt CosmoMC} package}.}. As discussed in Sec.~\ref{sec:NC}, we need this filter noise estimate to define the survey completeness function, a crucial input for making the connection between theory and the edited observations.  The \Planck collaboration has therefore already carried out many critical steps of the algorithm outlined in \sec{sec:topo_algo} and these available data products facilitate the topographic $y$-map analysis.

We now provide explicit details of how we use these data products to evaluate the remaining steps of the analysis, which essentially include slicing the $y$-map into uRC and RC by masking clusters detected above certain SNR threshold, the power spectrum analysis on the respective slices and estimation of errors on the measured power spectra. En route we carry out essential null tests, paving the way for the final model parameter analysis which we discuss in detail in Sec.~\ref{sec:mla}.

\begin{figure}
\centering
\includegraphics[width=0.98\columnwidth]{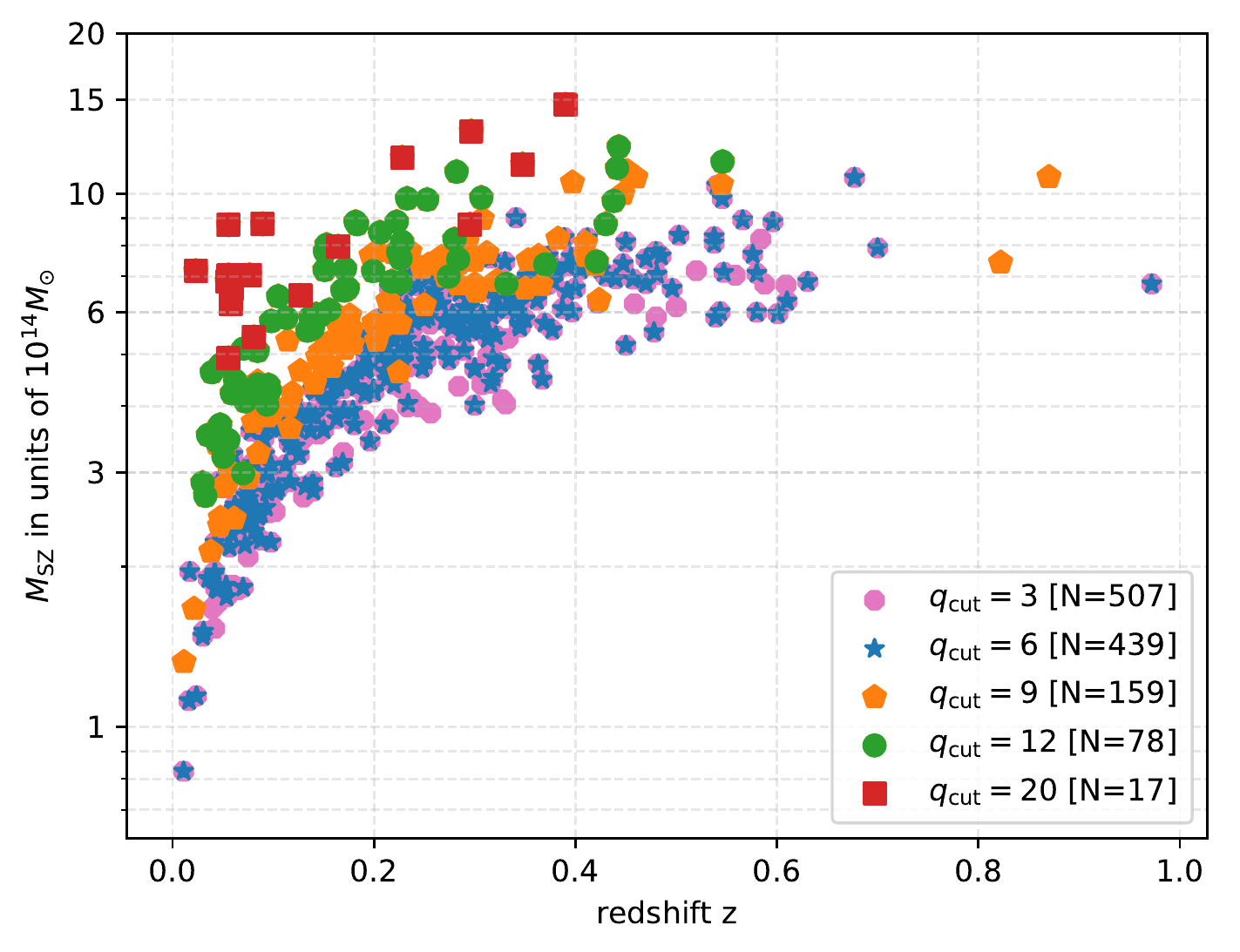}
\caption{Distribution of clusters in the $M-z$ plane for different $\qcut$. All clusters with $q\geq 6$ are included in the case $q\geq 3$. Note that higher $q$ typically corresponds to more massive clusters.}
\label{fig:M_z_qcut}
\end{figure}

\vspace{-3mm}
\subsubsection{Masking the giants}
We begin by reiterating that we refer to components of the $y$-map below the preset $q_{\rm cut}$ as uRC and the component of the $y$-map above $q_{\rm cut}$ as RC. We slice the $y$-map using a variety of SNR thresholds, specifically: $q_{\rm cut} \in [6,9,12,20]$. Figure~\ref{fig:M_z_qcut} depicts the distribution of the clusters detected by \Planck in the $M-z$ plane. Note that higher $q_{\rm cut}$ is generally associated with clusters of higher mass and lower redshifts. Also note that since we work with the MMF3 clusters in the COSMO sample, for $q_{\rm cut}=6$, the corresponding RC $y$-map is composed of the same sub-sample of clusters used in the \Planck CNC analysis. We now begin by describing how we construct the cluster mask, that enables us to decompose the $y$-map into uRC and RC for the different values of $q_{\rm cut}$. This mask is central to performing the PS analysis on the uRC and RC components of the $y$-map (see Sec.~\ref{sec:ps_ana}).

The MMF3 sample of clusters is used for the CNC analysis \citep{planck_cnc_2014}. We extract the SZ mass $M_{500}$ and sky coordinate ($\ell, b, z$) from this catalogue. Note that it is possible to obtain an angular size estimate for the cluster from the MMF analysis, by specifically providing $\theta$ at which the SNR estimate for a cluster detection is maximal. However, owing to the lack of this information we resort to estimating the projected angular size of the clusters, using the $M_{500}-\theta_{500}$, given in Eq.~(9) of \cite{Planck2013SZ}. We set $b=0.2$ and also use the redshift of the cluster, which is mostly available as part of the catalogue. Note that the results do not depend on the choice of `$b$', as our results are not sensitive to the size of the cluster mask.

With the estimated $\theta_{500}$, we next construct a cluster mask by punching a hole of radius $5 \times \theta_{500}$ centered on the sky location of each resolved cluster\footnote{We have tested a few different mask radii, specifically $3 \times \theta_{500}$, $5 \times \theta_{500}$ and $10 \times \theta_{500}$, and these yield very similar power spectra for the respective components of the $y$-map. Since the filter used in the MMF analysis also cuts of the profile at  $5 \times \theta_{500}$, we shall use this for all our analyses.}. For clusters in the sample with unknown redshifts, we use the median value of $\theta_{500}$, estimated from the rest of the clusters. Since COMA and VIRGO  are particularly large, we further extend the corresponding masks by multiplying with 3 degree radius masks centered on their galactic coordinates [i.e., VIRGO: $\left(\ell,b\right) \equiv \left( 279.68^{\circ}, 74.46^{\circ}\right)$; COMA:  $\left(\ell,b\right) \equiv \left( 58.08^{\circ}, 87.96^{\circ}\right)$]. The net resultant mask is denoted by $\mathcal{M}_C$.

\vspace{-0mm}
\subsubsection{The \Planck Compton $y$-map power spectrum analysis}
\label{sec:ps_ana}
Given $\mathcal{M}_C$, we now have all the inputs necessary to carry out the envisioned topographic $y$-map analysis. For most parts, our PS analysis follows closely the analysis carried out in \cite{Planck2016ymap}; however, we use an independent pipeline. Specifically in all our analysis, we cross correlate half mission 1 (HM1) NILC maps with half mission 2 (HM2) MILCA maps to estimate the $yy$-power spectrum\footnote{We also evaluate the $yy$-power spectra from NILC and MILCA maps by cross correlating ${\rm HM1} \times {\rm HM2}$, finding consistent results.}. To ease comparison with the \Planck \yy, we use the same multipole binning as provided in Table~12 of \citet{Planck2016ymap}.  
We use the \fskygal  mask for masking the dominant galactic contamination and the extended point source mask with \fskyps for removing contamination due infrared red and radio sources. We work with the union of these two masks which is apodized using a cosine-square profile with tapering carried out over a distance of \maskapo~ to reduce ringing near the sharp mask boundaries. The apodized union mask has an effective sky fraction \fskytotapo and will be denoted as $\mathcal{M}_G$.

Given the masks $\mathcal{M}_G$ and $\mathcal{M}_C$ we can now carry out three types of analyses. Estimating the power spectrum from the $y$-map masked with $\mathcal{M}_G$ results in the standard \Planck PS analysis. Note that in the language introduced in this work, this can also be interpreted as the uRC power spectrum in the limit of $q_{\rm cut} \rightarrow \infty$. The power spectrum estimated after masking the $y$-map with $\mathcal{M}_{\rm uRC} = \mathcal{M}_G \times \mathcal{M}_C$  yields \yy corresponding to uRC. Finally, the power spectrum estimated after masking the $y$-map with $\mathcal{M}_{\rm RC} =\mathcal{M}_G \times (1-\mathcal{M}_C)$  yields \yy corresponding to RC. Note that the apodization procedure on each of these masks is carried out after constructing the binary versions of the respective union masks. This apodization detail is particularly important for analysis carried out with $\mathcal{M}_{\rm RC}$, which is a particularly aggressive mask (see \fig{fig:masks} for $\mathcal{M}_{\rm RC}$ constructed assuming $q_{\rm cut}=6$). The cluster masks $\mathcal{M}_C$ are constructed for different values of $q_{\rm cut}$, and the evolution in the respective effective sky fractions (i.e. available sky fraction post apodization) associated with RC and uRC of the $y$-map are tabulated in Table~\ref{tab:fsky}. Note that the sky fractions associated with cluster masks, for different values of $q_{\rm cut}$ change by at most 1\%.
\begin{table}
\centering
\begin{tabular}{lccccc}
\toprule
 $q_{\rm cut}$ & 3  & 6 & 9 & 12 & 20  \\
\midrule
uRC (\%) &  34.36 & 34.46 & 34.90 & 35.07 & 35.26
\\
RC (\%)  & 1.22 & 1.10 & 0.60 & 0.41 & 0.20
\\
\bottomrule
\end{tabular}
\vspace{2mm}
\caption{This table summarizes the effective sky fraction (post apodization) associated with $\mathcal{M}_{\rm uRC}$ and $\mathcal{M}_{\rm RC}$ for different $q_{\rm cut}$. Note that these sky fractions include losses due to foreground contamination. For comparison note that the effective sky fraction associated with the foreground masks only is: $f_{\rm sky}^{\rm G}=35.44 \%$.}
\vspace{-3mm}
\label{tab:fsky}
\end{table}

We estimate the $y$-map power spectrum post masking using the MASTER algorithm \citep{Hivon2002}. In order to test robust working of our MASTER algorithm, we simulated a Gaussian $y$-map using a fiducial \yy and checked that the recovered PS is consistent with the input for both $\mathcal{M}_{\rm uRC}$ and $\mathcal{M}_{\rm RC}$ (see Appendix~\ref{app:test_master} for details). Since the Compton $y$-parameter field is highly non-Gaussian, there are some subtle issues arising from this and we discuss these nuances next.
However, before venturing into this discussion, we would like to draw attention to the fact that, while these nuances are important for sensible evaluation of the RC power spectra, they are not as critical to the estimation of the uRC power spectrum which will play a more important role in discussions of the subsequent sections (recall that it is the removal of high $q$ clusters, i.e., uRC of the $y$-map, which has its constraining power enhanced).

\vspace{-0mm}
\subsubsection{Nuances of estimating the MASTER-corrected PS} 
\label{sec:master_nuances}
The MASTER algorithm implicitly assumes that masked portion of the sky is composed of a field with statistical properties similar to that of the unmasked portion. Only under this assumption it corrects for the power lost in the masked sky fraction, while also accounting for mode coupling artefacts induced by masking. When estimating the MASTER-corrected, total Compton $y$-map power spectrum this assumption is valid, since it is reasonable to assume that the statistical properties of the $y$-field in the sky fraction lost to foreground contamination are identical to those in the sky fractions we observe\footnote{Strictly speaking this is not true, since varying foregrounds and measuring sensitivity amount to varying observing depths in different portions of the sky, which consequently must result in a modulation of the statistical properties of the $y$-map across these patches. In principle one would need to take this into account, but following all other analyses we ignore this detail.}.
However, when carving the $y$-map into portions that capture uRC and RC respectively, this assumption is strongly violated, since we are specifically masking certain peaks of the $y$-map, consequently altering the statistical properties of the field we observe. We can thus expect the default MASTER correction procedure to fail.
Indeed, the 1-point PDF of the total $y$-map, the uRC and the RC differ significantly, supporting the idea that masking Gaussianizes the map (see Appendix~\ref{app:1ppdf_study} for details). 
This can be verified by applying the default MASTER algorithm to the RC of the $y$-map (as we have done) and noted that this yields a \yy with an amplitude significantly larger than the total \yy. Clearly this is a consequence of the MASTER algorithm's implicit assumption that masked portions of the sky are covered with high SNR (massive) clusters as those included in the RC of the $y$-map. Therefore, applying the MASTER correction in its native form to estimate the PS of the RC is faulty. A similar issue arises when estimating the PS corresponding to the uRC; however, in this case the discrepancy is more subtle as the sky fraction is only changed by a few percent (see Table~\ref{tab:fsky}). To circumvent this issue we present two prescriptions for appropriately estimating the power spectrum corresponding to the RC and uRC of the $y$-map:

\vspace{1mm}
\noindent\textit{Method 1:} In this approach, we first estimate the MASTER-corrected PS using the default procedure and then correct the amplitude of the power spectrum by the factors: $A^{\rm RC} = f^{\rm RC}_{\rm sky}/f^G_{\rm sky}$ and  $A^{\rm uRC} = f^{\rm uRC}_{\rm sky}/f^G_{\rm sky}$ for the respective components. These correction factors can be interpreted as estimates of the mean sky fractions composed of the RC and uRC components of the $y$-map respectively. These correction factors are evaluated for different  values of $q_{\rm cut}$  and are summarized in Table.~\ref{tab:fsky}.

\vspace{1mm}
\noindent\textit{Method 2:} In this approach, the MASTER algorithm is evaluated assuming only the foregrounds mask $\mathcal{M}_G$, which then only corrects for the sky fraction lost to galactic and point source contamination. The additional area masked when using $\mathcal{M}_{\rm uRC}$  or $\mathcal{M}_{\rm RC}$ in this method can be thought of as an effective $M-z$ cut, bearing resemblance to the completeness modeling of the CNC analysis.

\vspace{-0mm}
\subsubsection{Robustness tests}
We now demonstrate that these two approaches result in consistent evaluation of the PS, except for differences on large angular scales, which can be expected. We also show that the sum of the RC and uRC power spectra returns the total \yy as one expects. Finally we also show that Gaussian errors on the respective power spectra, when added in quadrature, yield the error on the total \yy, modulo differences resulting from ignoring the noise contribution sourced by the cross correlation between RC and uRC of the $y$-map. We carry out identical tests on RC and uRC spectra derived from all values of $q_{\rm cut}$ used in our analysis, finding similar results. However, for brevity, we only present the specifics of tests carried out on spectral analysis performed with $q_{\rm cut}=6$. 
\begin{figure}
\centering
\includegraphics[width=1\columnwidth]{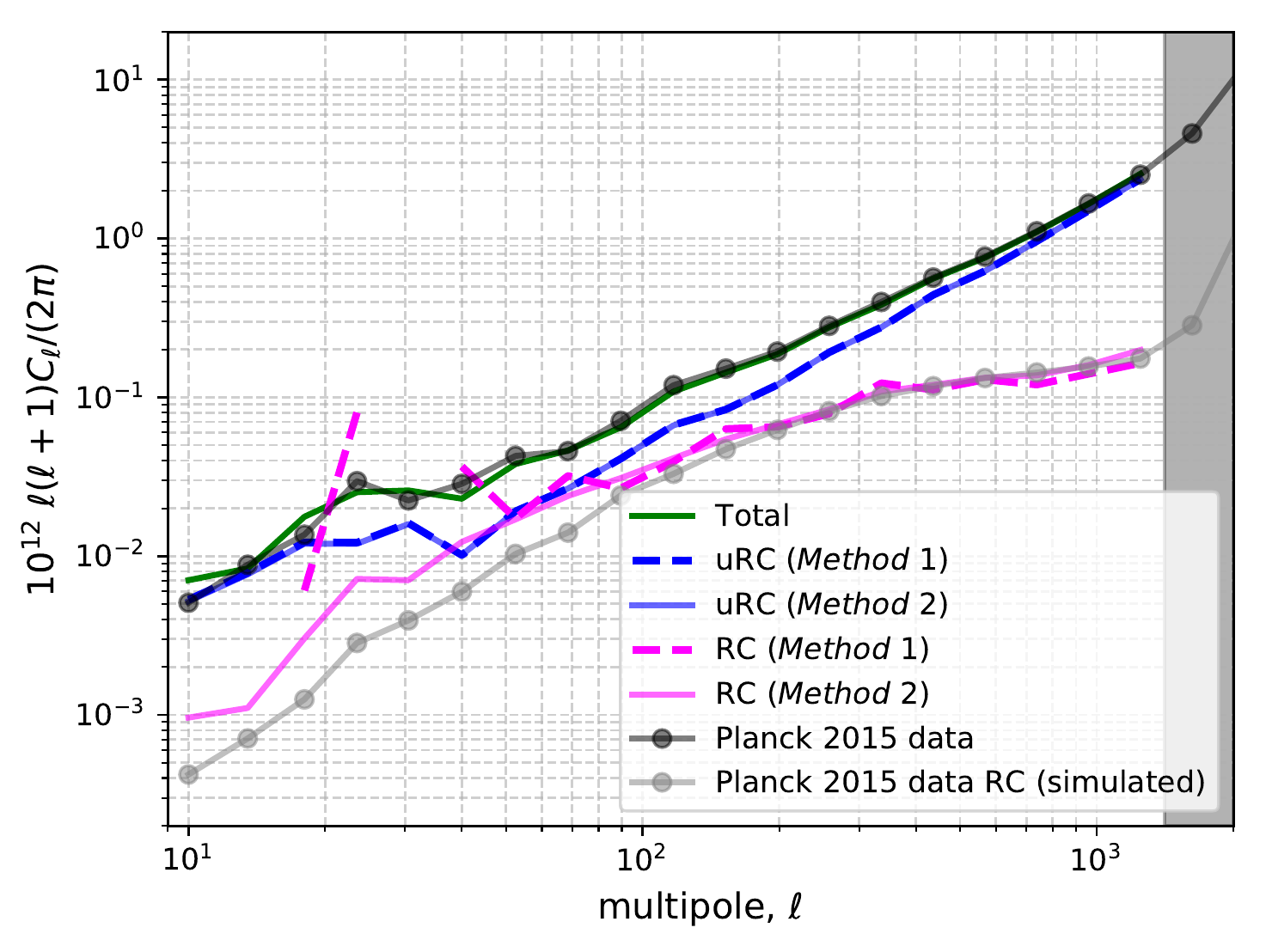}
\vspace{-2mm}
\caption{The green, blue and magenta curves show the $yy$-power spectrum corresponding to the total, uRC and RC of the \Planck $y$-map, evaluated for $q_{\rm cut}=6$. The black and gray curves depict the total and simulated RC component PS, as seen in Fig.~12 of  \citet{Planck2016ymap}.}
\vspace{-2mm}
\label{fig:yy_specs}
\end{figure}
\begin{figure}
\centering
\includegraphics[width=0.98\columnwidth]{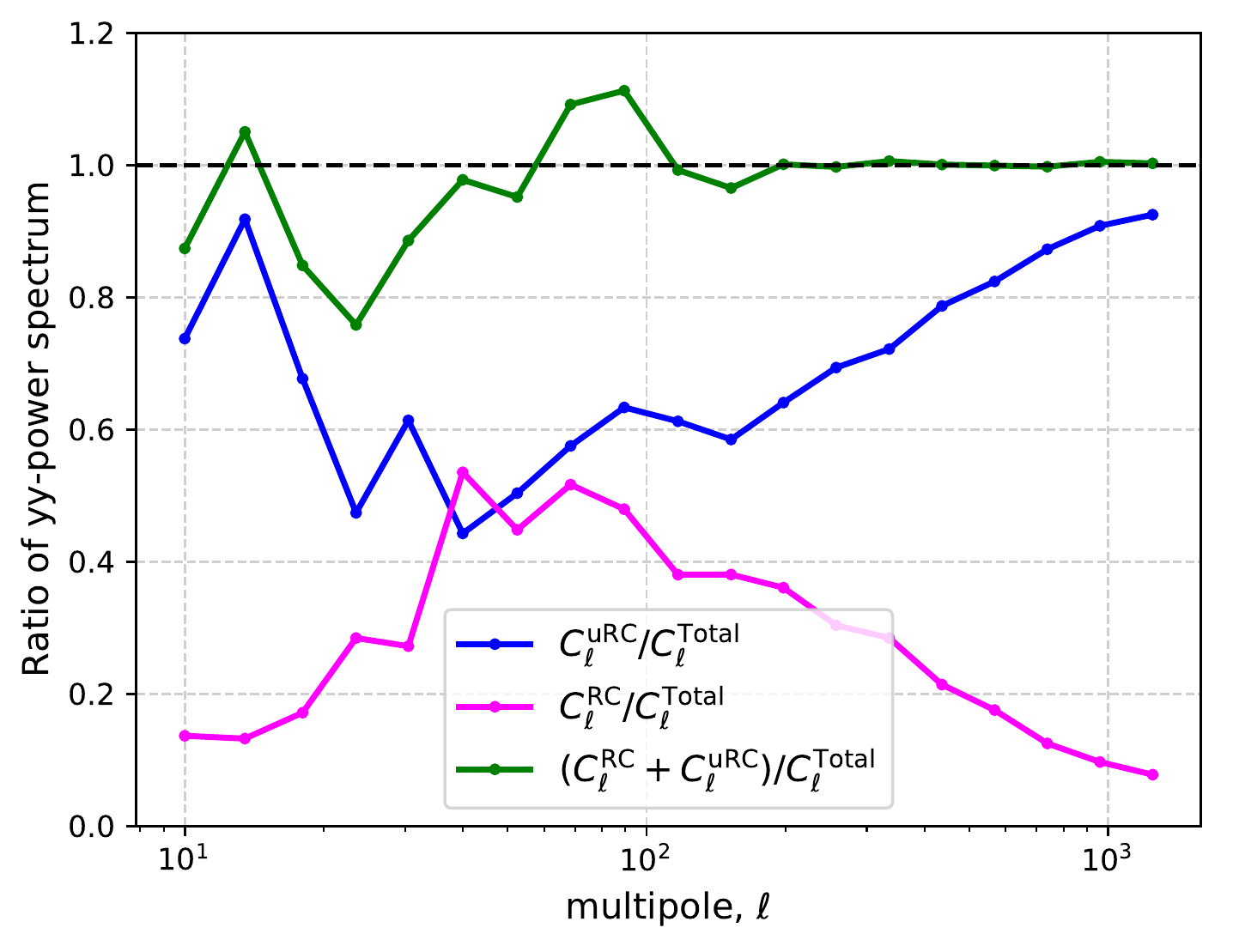}
\vspace{-2mm}
\caption{This figure depicts power spectrum contributions from the uRC and RC of the $y$-map relative to the total power spectrum, evaluated for $q_{\rm cut}=6$. Note that uRC and RC contributions add up to the total. uRC being composed of relatively less massive clusters, contribute dominantly at high multipoles while the contribution from the RC peaks at low multipoles owing to contributions dominantly originating from larger mass clusters.}
\vspace{-3mm}
\label{fig:yy_ps_null_test}
\end{figure}

\vspace{1mm}
\noindent\textit{Total power spectrum evaluation:} We first evaluate the SZ PS by using only the galactic and point source mask $\mathcal{M}_G$, appropriately apodized (\fskytotapo). We find good consistency between spectra derived using our analysis pipeline and those used in the \Planck 2015 analysis as seen in \fig{fig:yy_specs}. This serves as an additional test of our MASTER implementation, applied to real data.

\vspace{1mm}
\noindent\textit{uRC power spectrum evaluation:} We estimate the power spectrum corresponding to the uRC component of the $y$-map using both methods prescribed in \sec{sec:master_nuances}. The total $y$-map is masked with $\mathcal{M}_{\rm uRC} = \mathcal{M}_G \times \mathcal{M}_C$,  appropriately apodized (\fskydiffapo). With \textit{Method 1}, we evaluate the corrected PS using default MASTER algorithm and the amplitude of the resultant spectra is corrected by the factor \aurc. The  obtained \yy from both methods are found to be highly consistent (see \fig{fig:yy_specs}). Note that the primary effect of removing clusters detected above $q_{\rm cut}=6$ from the $y$-map is that of preferentially reducing power at high multipoles as compared to the total PS. This can also be seen in the top panel of \fig{fig:1halo2halo}, which shows the theoretical \yy for different values of $\qcut$.

\vspace{1mm}
\noindent\textit{RC power spectrum evaluation:} Similar to the uRC, we estimate the  the contribution to the power spectrum from the clusters in the MMF3 COSMO sample, forming the RC. In this case, the total $y$-map is masked with $\mathcal{M}_{\rm RC} = \mathcal{M}_G \times (1-\mathcal{M}_C)$,  appropriately apodized (\fskyresapo). The resultant spectra estimated using both methods are again found to be highly consistent for multipoles $\ell>50$ (see \fig{fig:yy_specs}). There are considerable differences in the power spectra recovered on large angular scales ($\ell \leq 50$) and this is primarily owing to the meagre sky coverage of $\mathcal{M}_{\rm RC}$ (see Fig.~\ref{fig:masks}). Note that our data derived estimates of the RC power spectrum shows good consistency with the COSMO sample PS depicted in Fig.~12 of  \cite{Planck2016ymap}, and reproduced here in \fig{fig:yy_specs}. 
Using both methods, we find slightly higher power in the RC power spectrum on large angular scales as compared to that presented by \Planck. It is important to appreciate the fact that, while the PS in the \Planck paper is estimated from a simulated $y$-map which assumes cluster profiles and injects estimated amplitudes of the cluster Compton $y$-parameter, the estimates presented here are direct measurements from the reconstructed $y$-map. 

\vspace{1mm}
\noindent\textit{Additional null test:} We expect that $C^{yy}_{\ell} \simeq C^{\rm uRC}_{\ell} +C^{\rm RC}_{\ell}$. We find that RC and uRC spectra estimated using both the methods satisfy this constraint. We evaluate the relative amplitudes $C^{\rm uRC}_{\ell} /C^{\rm Total}_{\ell}$ and $C^{\rm RC}_{\ell} /C^{\rm Total}_{\ell}$ and show that, modulo expected variance, they sum to unity in each multipole bin as depicted in \fig{fig:yy_ps_null_test}. Presenting this test in terms of relative amplitude allows us to highlight the minor differences in the total spectra evaluated directly and by summing the uRC and RC contributions to the spectrum. Note that on large angular scales $\ell\lesssim100$ the differences can be up to $\simeq 20 \%$; however, for multipoles $\ell>100$ the consistency is extremely good. 

\vspace{1mm}
Having demonstrated that the power spectra evaluated using the two methods are consistent with each other and owing to the more desirable behaviour of the RC spectra evaluated using \textit{Method~2}, particularly on large angular scales, for the rest of the analysis we work with spectra evaluated using this method.

\vspace{-3mm}
\subsubsection{Map-based Gaussian error estimation}
\label{sec:error_est}

\begin{figure}
\centering
\includegraphics[width=0.98\columnwidth]{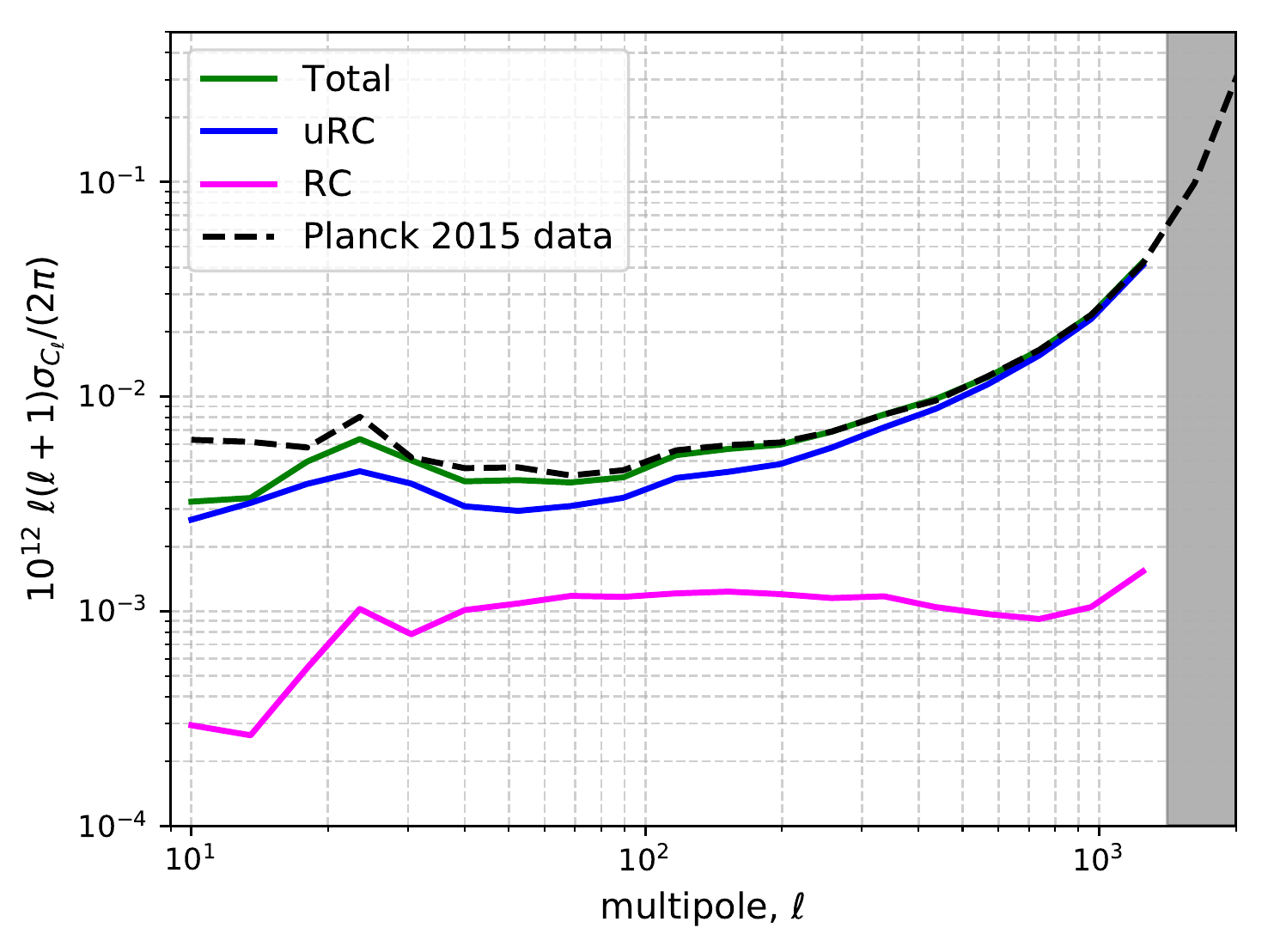}
\caption{Gaussian error estimates on the power spectrum for total, uRC and RC components of the $y$-map evaluated for $q_{\rm cut}=6$. The black dashed lines depicts the error on the power spectrum evaluated by the \Planck collaboration. The difference in snoise estimate on large scales is due to non-inclusion of striping noise in our analysis.}
\label{fig:err_spec}
\end{figure}
The Gaussian part of the error on the measured \yy is estimated directly from data and here we discuss the details of its evaluation. It can be shown that the error on the power spectrum estimated from two statistically independent measurements of the $y$-map is given by the following equation,
\beq
\sigma^2_{C_{\ell}^{yy}}
=\frac{2 (C_{\ell}^{yy})^2 
+ C_{\ell}^{yy} \left(C_{\ell}^{N_1} +C_{\ell}^{N_2} \right)
+ C_{\ell}^{N_1} C_{\ell}^{N_2}}{(2 \ell+1) \Delta \ell_{\rm bin} f_{\rm sky}^{\rm eff}} \,, 
\label{eq:spec_noise_est}
\eeq
where $C_{\ell}^{N_1}$ and $C_{\ell}^{N_2}$ denote the measurement noise power spectra of the two statistically independent measurements and all other symbols have their usual meaning. In practice the error on the measured power spectrum is computed using \citep{Tristram2005}
\beq
\hat{\sigma}^2_{C_{\ell}^{yy}}=\frac{(\hat{C}_{\ell}^{{\rm HM1} \times {\rm HM2}})^2 + \hat{C}_{\ell}^{{\rm HM1} \times {\rm HM1}} \hat{C}_{\ell}^{{\rm HM2} \times {\rm HM2}}}{(2 \ell+1) \Delta \ell_{\rm bin} f^{\rm eff}_{\rm sky}} \,.
\eeq
We compare our noise estimates on the total \yy with those from \Planck and find fully compatible results as shown in \fig{fig:err_spec} (compare dashed line with solid green line). 
We suspect that the small differences between our noise estimates and those from \Planck at low multipoles are due to neglecting excess errors due to striping in the $y$-maps in our analysis\footnote{The \Planck collaboration specifically mention that they add this to their noise estimates, but neither this excess noise estimate nor the procedure to estimate it are discussed.}. 
This detail makes little difference for the total PS analysis since in this case the noise at low multipoles is dominated by non-Gaussian terms, derived from theory. However on using a low $q_{\rm cut}$ this error contribution can become important to the uRC component at low-$\ell$ and we revisit this detail in \sec{sec:2_halo}. 
We use an identical procedure to estimate errors on the power spectrum corresponding to the RC and uRC of the $y$-map. Adding the noise estimates on the power spectrum of uRC and RC components in quadrature returns nearly the noise on the total \yy, but with a small expected deficit. This deficit in noise power is the error on the cross correlation between the uRC and RC  $\propto 2 C_{\ell}^{\rm uRC} C_{\ell}^{\rm RC}$.

\vspace{-3mm}
\subsubsection{Total error estimation}
\label{sec:total_error_est}

\begin{figure}
\centering
\includegraphics[width=0.98\columnwidth]{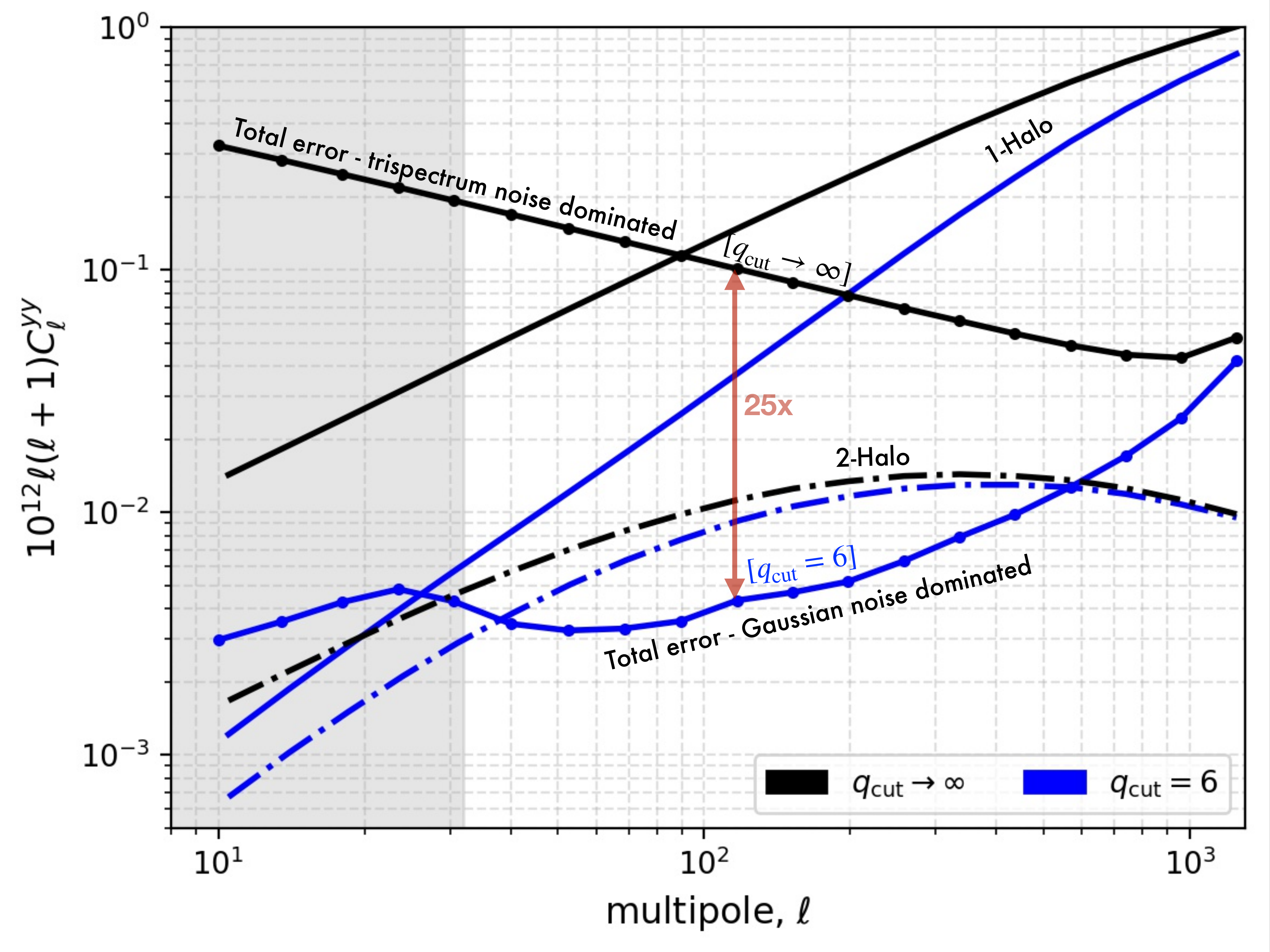}
\caption{The $yy$-PS and their statistics for total y-map and for uRC corresponding to $q_{\rm cut}=6$. The total error depicted here includes the Gaussian as well as the trispectrum noise evaluated for \Planck binning. Note that for total, the 2-halo contribution is statistically un-detectable, whereas for uRC the 2-halo contribution is above the total error at $\ell \simeq 40-500$.}
\label{fig:qcut_err_spec}
\end{figure}

As noted before, the Compton-$y$ field is highly non-Gaussian and consequently the errors on the PS receive significant contribution from the trispectrum. This non-Gaussian covariance is estimated following the procedure detailed in \sec{sec:tri_est}. We draw attention to \fig{fig:yy_err_qcut}, which shows the evolution of the diagonal of the trispectrum covariance for different $q_{\rm cut}$. In particular for the uRC note that the trispectrum errors reduces on lowering the cluster detection SNR threshold, even becoming comparable to the Gaussian noise variance for sufficiently low values of $q_{\rm cut}$.

The total error covariance is now obtained as a sum of the Gaussian covariance estimated from data and the trispectrum covariance which is calculated duly accounting for $f_{\rm sky}$.
In Fig.~\ref{fig:qcut_err_spec}, we show diagonal of the error covariance along with a decomposition of the different contributions to the $yy$ power spectrum. Firstly, we note that for the total $y$-map the 2-halo contribution is burried under statistical uncertainties when the trispectrum noise is properly accounted for and the 2-halo term can be completely ignored. In the original \Planck analysis \citep{Planck2016ymap}, the trispectrum noise was omitted, thus necessitating the inclusion of the 2-halo terms, although insignificant at the corrected error budget. 
More interestingly, on masking $q\geq 6$ clusters in the \Planck $y$-map we note that trispectrum is suppressed so much so that the total error is nearly equal to the Gaussian noise alone (cf. \fig{fig:qcut_err_spec}). Although the 2-halo contribution is mildly reduced, it is still above the total expected statistical uncertainty at $\ell \simeq 40-500$. The corresponding uRC power spectra are therefore expected to be sensitive to the 2-halo contribution as is seen in \fig{fig:qcut_err_spec}.
Note, however, that the Gaussian variance estimated directly from the map is strongly dominated by foreground residuals. The Gaussian error contribution is thus still significantly above the expected theoretical Gaussian cosmic variance terms (cf. Fig.~\ref{fig:yy_err_qcut}).

\begin{table}
\begin{centering}
\begin{tabular}{lccc}
\toprule
{} &  1 + 2 halo &  1-halo &  2-halo \\
\midrule
Total ($q_{\rm cut} \rightarrow \infty) \;\;[T_{\ell\ell}]$ 
&        36 (3.7) &    36 (3.4) &     0.7 (0.3) \\
$q_{\rm cut}=6 \qquad\qquad\;[T_{\ell\ell}]$  
&        66 (8.6) &    62 (5.1) &     6.1 (3.7) \\
\hline
Total ($q_{\rm cut} \rightarrow \infty) \;\;[T_{\ell\ell'}]$ 
&        25 (3.5) &    24 (3.3) &     0.3 (0.2) \\
$q_{\rm cut}=6 \qquad\qquad\;[T_{\ell\ell'}]$  
&        47 (8.3) &    45 (5.1) &     4.6 (3.4) \\
\bottomrule
\end{tabular}
\par\end{centering}
\caption{Approximate SNR of the different components of the fiducial power spectra, estimated assuming the total noise for each value of $q_{\rm cut}$ (cf. see \fig{fig:err_spec}). Two treatments of the trispectrum contributions are used. The numbers in brackets are Fisher estimates of the SNR after marginalizing over the foreground templates for CIB, RS and IR.}
\vspace{-3mm}
\label{tab:snrs}
\end{table}
To quantify this aspect further, we approximately estimate the SNR with which the different components of a  fiducial \yy power spectrum are expected to be measured. The SNR is simply evaluated using the following expression,
\begin{align}
{\rm SNR} = \left[ C^{ yy}_{\ell} M^{-1}_{\ell \ell'} C^{yy}_{\ell'} \right]^{1/2} \,,
\end{align}
where $M_{\ell \ell'}$ denotes the total noise covariance. Note that this expression identically matches the Fisher estimate for a single component analysis. 
In our approximate treatment, $M_{\ell \ell'}$ accounts for the trispectrum error contributions, where we consider the cases with and without off-diagonals. 

The SNR estimates with and without foreground marginalization are reported in Table~\ref{tab:snrs}. 
As expected, for the total $y$-map, the 1-halo and 1+2-halo description of the $yy$-PS are indistinguishable. On masking cluster $q\geq6$, the SNR improves by a factor of $\simeq 2$, suggesting an enhancement of the parameter constraint by a similar factor. We further see that now the 2-halo contribution becomes statistically important, suggesting a $\simeq 6.1\sigma$ detection could be possible with \Planck data. On including foreground marginalization the estimated statistical significance of detection expectedly drops but all the above discussed trends are retained. The foreground-marginalised estimates continue to suggest a marginal detection of the 2-halo at $\simeq 3.7\sigma$. We also note that the 2-halo contribution is less affected by foregrounds as the SNR is reduced only by a factor of $\simeq 2$ as opposed to the 1-halo term whose detection significance is reduced by a factor of $\simeq 10$.
Finally, off-diagonal contributions make a $\simeq 30\%-40\%$ difference in the expected SNR for our analysis without foregrounds. When foregrounds are included, the difference is minute.
We discuss the implications of these estimates in \sec{sec:mla}.

\vspace{-3mm}
\subsubsection{Final power spectra}
\label{sec:power_spectra_uRC_RC}
\begin{figure}
\centering
\includegraphics[width=1\columnwidth]{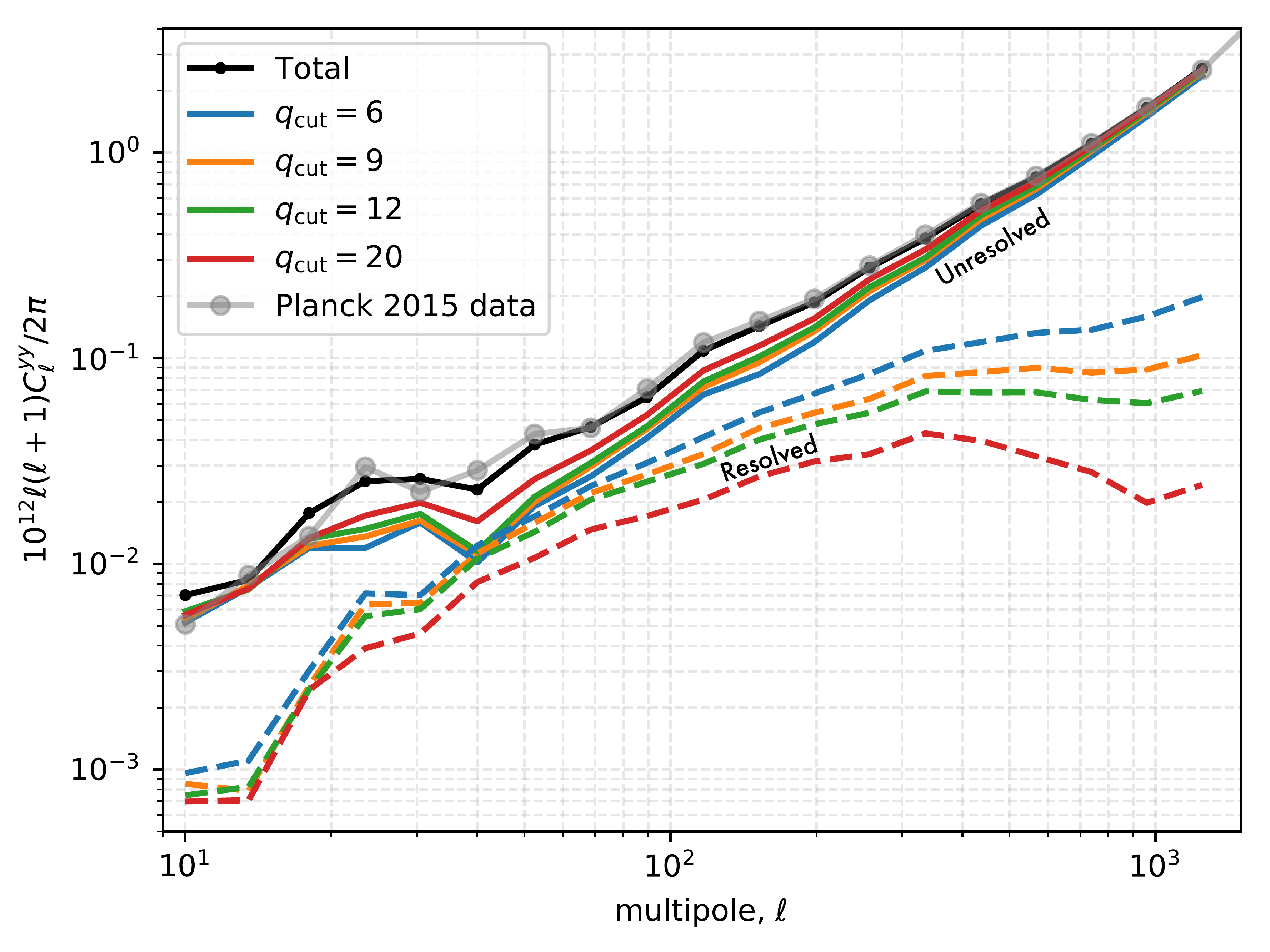}
\vspace{-2mm}
\caption{Final results for the obtained PS with completeness modeling. The solid lines show the uRC power spectra and dashed lines depict the RC power spectra, with different colors denoting the corresponding $q_{\rm cut}$. These include foregrounds that are marginalized over in our analysis.
The data for this figure with be provided at \TopoSZ.}
\vspace{-0mm}
\label{fig:all_yy_specs}
\end{figure}

In Fig.~\ref{fig:all_yy_specs}, we describe the dependence of the final power spectra on $\qcut$, both for the uRC (solid lines) and RC (dashed).
These are the sum of all contributions in the map, containing the $y$-distortion signal and foregrounds that will be marginalized over in Sect.~\ref{sec:mla}.
At small scales ($\ell\gtrsim 10^3$), instrumental noise starts dominating for \Planck, such that we do not present this part in more detail. At $\ell\lesssim 10^3$, instrumental noise is mostly subdominant.

The amplitude of the uRC decreases when lowering $\qcut$, as less power from SZ clusters remains in the map. The reduction is most noticeable at $\ell \simeq 50-100$, suggesting that at both $\ell\lesssim 20-30$ and $\ell \gtrsim 200$ foregrounds could be strongly contributing.
Foregrounds can potentially be further reduced in the $y$-map by using constrained-ILC methods \citep[e.g.,][]{Remazeilles2011cILC, remazeilles2020rsz_map,rotti2020milc}, however, a more detailed discussion is beyond the scope of this work, also because we want to stay as close as possible to the standard \Planck $yy$-PS analysis.
For the RC of the map, we observe the opposite trend with $\qcut$, as expected. The shape and level of the RC-PS arguably  resembles the theoretical predictions shown in the lower panel of Fig.~\ref{fig:1halo2halo}, building confidence in the proposed approach.

\vspace{-0mm}
\section{Maximum likelihood analysis and results}
\label{sec:mla}
\label{sec:results}
Using the RC and uRC power spectra estimated for different values of $\qcut$ complemented with the corresponding estimates of Gaussian and non-Gaussian noise as detailed in \sec{sec:tri_est} and \sec{sec:error_est} respectively, we now carry out the likelihood analysis on each of the derived datasets. 
For this analysis we closely follow the procedures detailed in\footnote{We explored the possibility of fixing the covariance matrix and combining with an iterative approach, as suggested in \cite{Makiya:2019lvm}. While for $\qcut=6$ this approximation did not affect the results, for the total PS analysis, this increased the final error on $b$ by a factor of $\simeq 4$, introducing heavy wings to the posteriors of $F_{\rm SZ}$ (see Appendix~\ref{app:test_cov} for discussion).} \cite{Bolliet2017}. 
There it was shown that the SZ power spectrum is mainly sensitive to the parameter combination
\begin{align}
F_\mathrm{SZ}=\left( \frac{\sigma_8}{0.8} \right) \left(\frac{\Omega_{\rm m}}{0.3} \right)^{0.35} \left( \frac{B}{1.25} \right)^{-0.35} h_{70}^{-0.2}. \label{eq:ftsz}
\end{align}
The scaling with the Hubble constant slightly depends on the choice of the pressure profile. Here, we quote the one for the \cite{Arnaud2010} pressure profile parameterization. Written in this way, $F_{\rm SZ}$ is expected to be close to unity for standard assumptions and cosmological model. Since strictly speaking one cannot expect that with the new completeness modeling or inclusion of the 2-halo terms the parameter combination in Eq.~\eqref{eq:ftsz} is exactly valid, we consider the limitations of this treatment in Sect.~\ref{sec:2_halo}; however, the main conclusions are unaffected by this choice.

In our analysis we vary all relevant cosmological parameters, as well as the mass bias (the neutrino masses are kept fixed to our fiducial value), and obtain constraints on the parameter combination of Eq. \eqref{eq:ftsz}. 
The full trispectrum is taken into account in the likelihood.
We keep the amplitude of the correlated noise fixed to $A_\mathrm{CN}=0.903$ and eventually marginalise over the other three amplitudes of foreground residuals corresponding to infrared source (IR), cosmic infrared background (CIB) and radio sources (RS): $A_\mathrm{IR}$, $A_\mathrm{CIB}$ and $A_\mathrm{RS}$. For these foregrounds we use the same power spectrum templates as those used in the original \Planck analyses [\citet{Planck2013ymap} and \citet{Planck2016ymap}] and these are available from Table~3 of \cite{Bolliet2017}. For CIB in particular, we estimate the CIB power spectra using the \Planck CIB maps \citep{Planck2016GNILC} to confirm that there are no significant deviations in the shape of the template for the different masks used in the uRC analysis, hence justifying the continued use of the original CIB power spectrum  template.

The uniform priors we used for the varying parameters of the MCMC are reported in Table \ref{tab:priors}. For the sampling we ran up-to-date versions of \verb|Montepython| \citep{brinckmann2018montepython} and \verb|cobaya| \citep{Torrado:2020dgo}, which gave identical results. To compute posterior probability distributions and contours we use \verb|GetDist| \citep{Lewis:2019xzd}. For the theory predictions of the SZ power spectrum we use \verb|CLASS_SZ| \citep{Bolliet2017, Bolliet:2019zuz}.  
Using this likelihood setup, we now focus on discussing the key results of our analysis in the following sections.


\vspace{-0mm}
\subsection{A solution to the low $\ell$ degeneracy between foreground residuals and SZ?}
\label{sec:frg_param_evol}
\begin{figure}
\centering
\includegraphics[ clip,width=1\columnwidth]{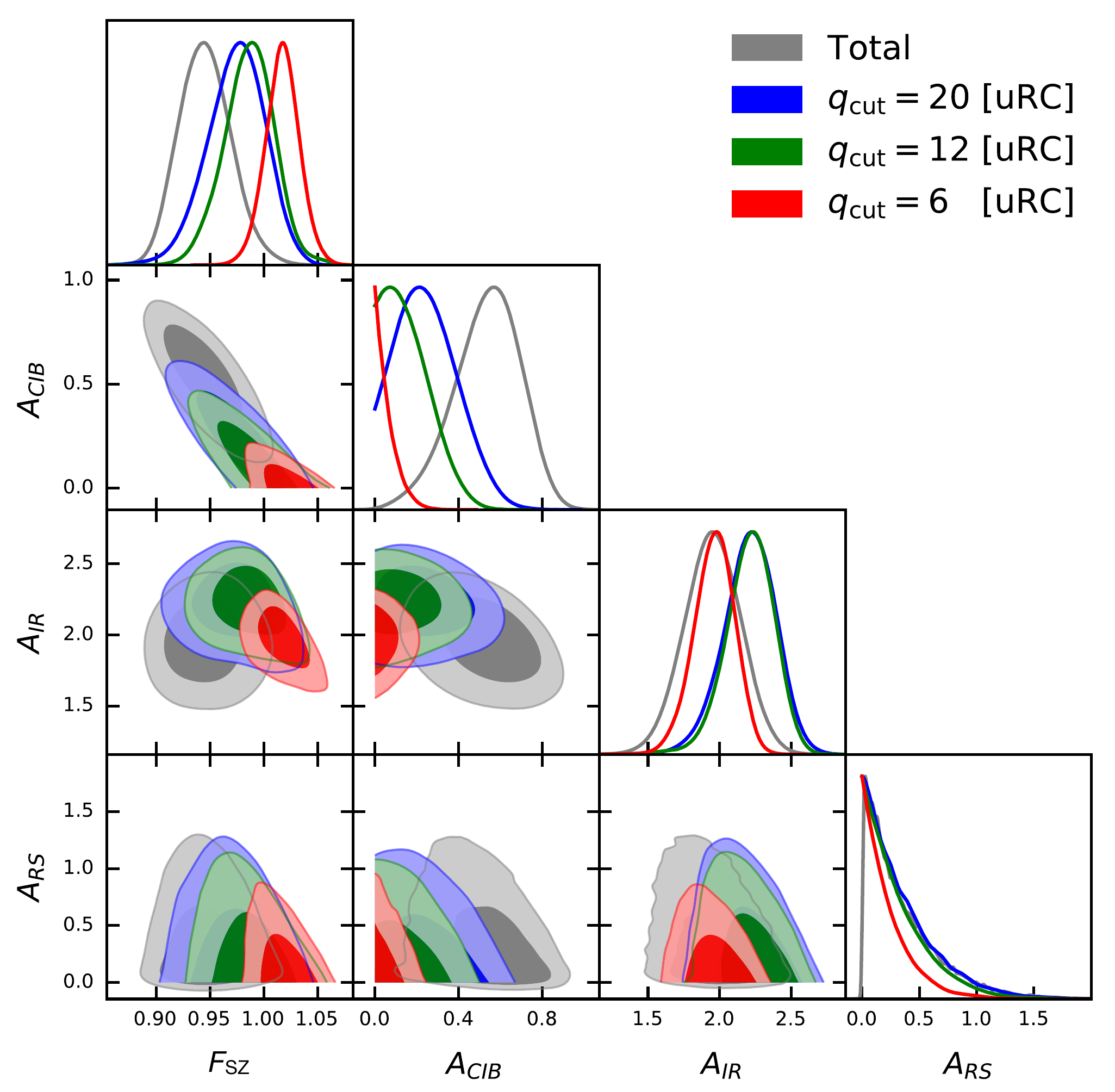}
\vspace{-0mm}
\caption{Parameter constraints from the measured resolved SZ power spectrum for \Planck. The CIB amplitude is highly degenerate with the SZ power spectral amplitude. This decoupling can be understood from the distinct CIB and resolved cluster SZ power spectral shapes.
}
\vspace{-0mm}
\label{fig:cosmo_constraint_unresolved}
\end{figure}

We begin by highlighting that in this section we only use the 1-halo contribution to model the $yy$-PS.
To understand the main effects of the novel $\qcut$-modeling, we present the $68\%$ and $95\%$ Confidence Level (CL) parameter contours for various values of $\qcut$ in Fig.~\ref{fig:cosmo_constraint_unresolved}. 
The grey contours are the results from the total $yy$-PS analysis, i.e., without using any completeness modelling in the SZ-PS, thus resembling most closely the standard \Planck PS analysis.

We note that here we chose to not use a prior bound for the amplitude of total foreground residual, as usually imposed \citep[e.g., see Eq.~(14) of][]{Bolliet2017}. This prior condition ensures that the SZ contribution to the PS is always larger than the PS of projected profiles of resolved clusters from the \Planck catalogue. 
Now this aspect becomes more important, since here we introduce a new masking strategy and perform analyses where some of these clusters are indeed masked, generally invalidating the condition. Hence, for consistency, here we avoid any explicit condition on the minimal amplitude that the SZ power spectrum of a masked map should have. This allow us to quantitatively compare the results for different $q_\mathrm{cut}$ with the \Planck total SZ PS analysis.

For the total PS analysis, we can also observe a clear anti-correlation between the cosmological parameter combination and the amplitude of CIB (see top panel of \ref{fig:cosmo_constraint_unresolved}). This arises because without this prior bound, due to the large trispectrum at low $\ell$ (see Fig.~\ref{fig:yy_err_qcut}), a significant fraction of the $yy$-PS can be explained by a sum of foreground residuals (mainly CIB and IR) with a reduced \yy amplitude. But as we shall see now, masking the heaviest clusters from the $y$-map alleviates this problem.

As we decrease the cluster SNR cut from $q_\mathrm{cut}=20$ (i.e., masking $17$ halos see Fig \ref{fig:M_z_qcut}), to $q_\mathrm{cut}=12$ (i.e., masking $78$ halos) and $q_\mathrm{cut}=6$ (i.e., masking $439$ halos), the surfaces of the 2d contours tend to decrease in size for all pairs of parameters (see Fig.~\ref{fig:cosmo_constraint_unresolved}). This is indeed expected, since masking the heavy clusters generally yields a gain in SNR for the SZ power spectrum as was discussed in Sec.~\ref{sec:total_error_est}.
%
%
Specifically, the constraint on $F_{\rm SZ}$ improves from  $F_{\rm SZ}= 0.935\pm 0.026$ for the total to $F_{\rm SZ}=1.0170\pm0.0160$ for $\qcut=6$.
Most strikingly, as we mask more clusters, the CIB component is driven to a nearly negligible level, where, e.g., for ${q_\mathrm{cut}=6}$ we find $A_\mathrm{CIB}<0.2$ (95\%CL), as opposed to $A_\mathrm{CIB}<0.9$ (95\%CL) for the total PS analysis. This means that the PS of the masked $y$-map, which is slightly steeper than the total (see Fig.~\ref{fig:all_yy_specs}), cannot accommodate a large CIB. Meanwhile, the posterior of the IR amplitude remains roughly the same and the RS amplitude is slightly less likely to take on large values as we mask more cluster. 

Since the degeneracy between CIB and SZ is an anti-correlation (i.e., more CIB requires less SZ and vice versa), having the CIB amplitude driven to smaller values when we mask heavy clusters implies an enhanced amplitude for the SZ-PS. 
As we mentioned above, the amplitude of SZ is determined by $F_{\rm SZ}$, and therefore the central value for this parameter is seen to shift upwards (see bottom left corner of Fig.~\ref{fig:cosmo_constraint_unresolved}), towards values more consistent with standard cosmology and assumptions (for which this combination approaches unity). 
Our findings thus suggests that clusters detected with a large SNR or correspondingly clusters of high mass have a larger CIB contamination associated with them. This observation is consistent with results of \citet{planck2016_cib_tsz}, where \Planck team reported a high significance $6\sigma$ detection of the CIB-tSZ correlation by performing a stacking analysis of confirmed clusters. In their analysis it was further noted that, the detection significance drops to $3 \sigma$ on using the complete $y$-map, suggesting that uRC component primarily adds noise, resulting in the reduction of the detection significance of the CIB-tSZ correlation. This is also in line with the more recent cross-correlation study \citep[Fig.~5 of][]{Chiang2020}, and thus highlights the potential of the proposed completeness modelling as a diagnostic tool.
Indeed, this also affects the constraints on the mass bias, as we discuss next.  

\vspace{-0mm}
\subsection{A consistent measurement of the mass bias in $\Lambda$CDM?}
\label{sec:mass_bias}
Since the $y$-map PS analysis provides a constraint on the combination in Eq.~\eqref{eq:ftsz} between cosmological parameters ($\sigma_8$ and $\Omega_\mathrm{m}$) and pressure profile parameters (the mass bias $B$), it is difficult to draw conclusions on the favored cosmological model, or favored pressure profile model unless we bring in extra information. For instance, one can assume a tight prior on the mass bias, motivated by hydrodynamical simulations results, and derive constraints on the cosmological parameters. This is what was originally done in \cite{Planck2013ymap} and \cite{Planck2016ymap} where, assuming a standard mass bias of $b=0.2$, the final constraint on matter clustering was $\sigma_8=0.77\pm 0.02$ (68\% CL). 
This is 2 standard deviations lower than the constraints from primary CMB, namely $\sigma_8=0.811\pm 0.006$ (68\% CL) \citep{Aghanim:2018eyx}, signifying one of the current cosmological tensions.
Alternatively, one can  assume a cosmological model and obtain a measurement of the mass bias within this model. This can be achieved by jointly analysing primary CMB data with the $y$-map PS, or, more simply, by combining the measurement of the combination $\sigma_8 \Omega_{\rm m}^{0.35} h_{70}^{-0.2}$ from the primary CMB with the constraint on the combination of Eq.~\eqref{eq:ftsz} to deduce the value of the bias \citep[see ][]{Bolliet2017}.

The resulting value of the mass bias that was obtained in previous works \citep{Bolliet2017,Bolliet:2019zuz} can be summarized by $b=0.40\pm 0.05$ (68\%CL), using the \Planck 2015 total $y$-map power spectrum and primary CMB data. 
Note that this is nearly the same as the constraint obtained using the \Planck CNC (jointly with primary CMB data), namely $b=0.42\pm0.04$ (68\%~CL) \citep{Ade:2015fva}, or the results from the re-analysis of \citet{Salvati_2019} who found $b=0.38\pm0.05$ (68\%~CL). The \Planck $y$-map power spectrum analysis paper did not present a constrain on the mass bias parameter.

In what follows, we adopt the second method: we analyse the \Planck 2018 $\Lambda$CDM chains\footnote{plikHM\_TTTEEE\_lowl\_lowE} to obtain the constraint on cosmological parameters:
\begin{equation}
    F^{\rm CMB}_{\rm SZ}=\left(\frac{\sigma_8}{0.8}\right)\left(\frac{\Omega_{\rm m}}{0.35}\right)^{^{0.35}}h_{70}^{^{-0.2}} = 1.042\pm 0.018\,\, (68\%~\mathrm{CL}).\label{eq:plc18cosmo}
\end{equation}
and then derive the constraints on the mass bias corresponding to the \Planck $\Lambda$CDM cosmological model, by combining Eq.~\eqref{eq:plc18cosmo} with the constraints on the combination of Eq.~\eqref{eq:ftsz} from our analyses of the $y$-map (i.e., the 1d posterior PDF in the top panel of Fig.~\ref{fig:cosmo_constraint_unresolved}). 

\begin{figure}
\centering
\includegraphics[width=0.8\columnwidth]{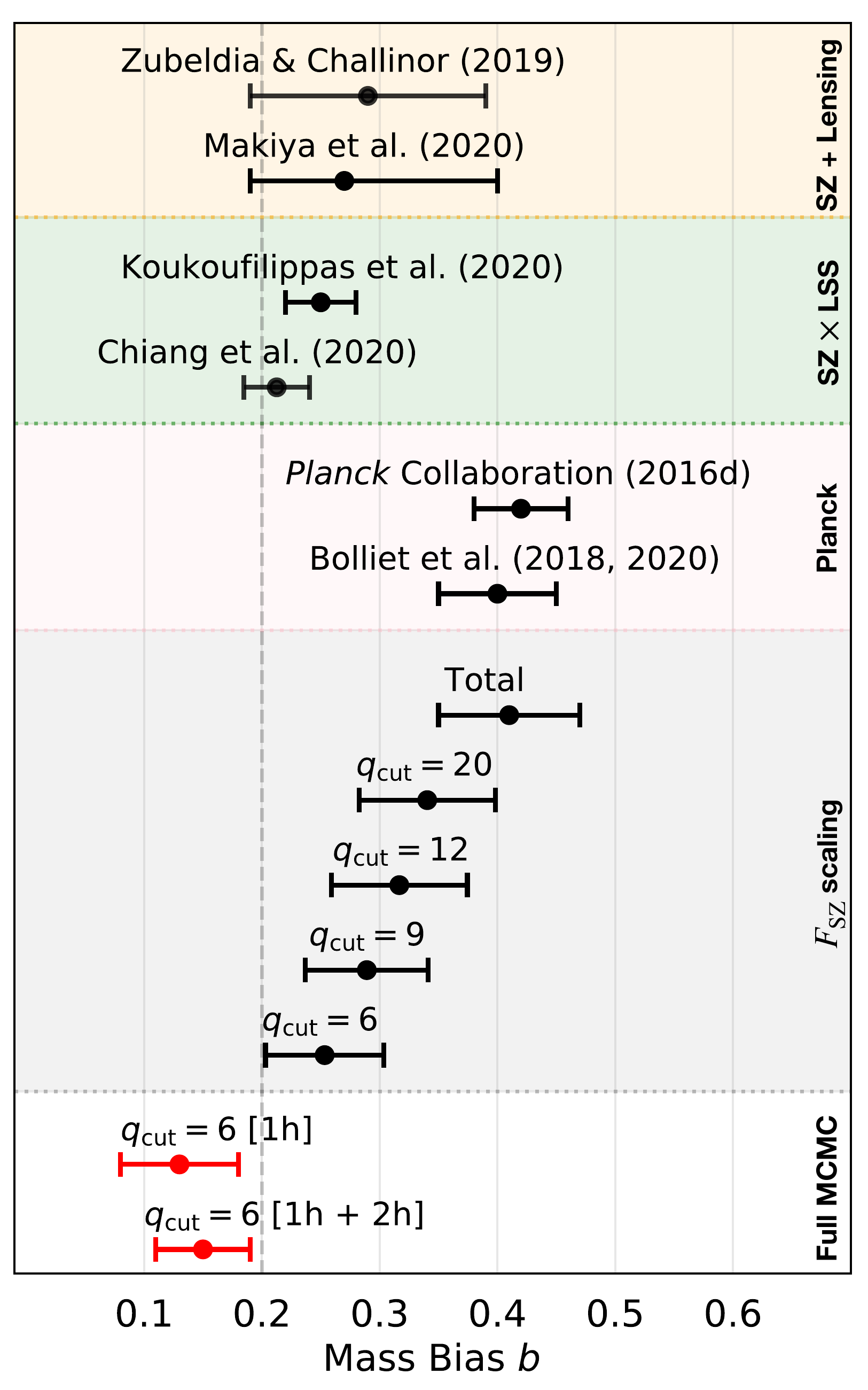}
\caption{Mass bias from our analyses with several values of the cluster SNR cut $q_{\rm cut}$, and  measurements from previous work for comparison. 
%
See Sect.~\ref{sec:mass_bias} for detailed discussion.
}
\vspace{-3mm}
\label{fig:bias_evol}
\end{figure}

Our results for several values of cluster SNR cut $q_\mathrm{cut}$ are shown on Figure  \ref{fig:bias_evol} and reported in Table~\ref{tab:biasconstraints}. 
For comparison, we also quote several other measurements of $b$ from the literature.
These are derived by combining SZ and CMB data from \Planck with lensing measurements \citep{Makiya:2019lvm, Zubeldia2019} and by cross-correlating with galaxy catalogs \citep{Koukoufilippas2020, Chiang2020}.
As the cluster SNR cut decreases and more clusters are masked, the posterior probability distribution of the mass bias shifts towards lower values of the mass bias $b$, which corresponds to a higher amplitude of SZ relative to CIB.  
Although for $F_{\rm SZ}$ we found error improvements with $\qcut$, these do not translate directly to $b$. 

For $q_\mathrm{cut}=20$, when only a few clusters are masked, we find a result close to the previously reported measurement, which lies in the upper part of the range suggested by hydrodynamical simulation and analytic calculations of non-thermal electron pressure with a typical value of $b\simeq 0.2\pm 0.1$ \citep[see, e.g.,][]{Nagai2007, Shaw2010SZ, Battaglia2012, Shi_2014, Shi_2015, Biffi2016}. For $q_\mathrm{cut}=6$, i.e., when we mask all the clusters of the \Planck COSMO sample, we find a mass bias of $b=0.25\pm0.05\,(68\%~\mathrm{CL})$,
consistent with theoretical expectations. 
Although other recent analyses \citep{Zubeldia2019, Makiya:2019lvm, Koukoufilippas2020, Chiang2020} have reported measurements of the mass bias consistent with the theory predictions, some of these works rely extensively on data from galaxy surveys.

It is also important to highlight the noticeable evolution of the central value in the mass bias when lowering the value of $\qcut$ (Fig.~\ref{fig:bias_evol}). This could be taken as a hint for mass-dependent bias in the sample of clusters, since for varying $\qcut$ the effective mass relevant to the PS reduces (see Fig.~\ref{fig:mean_mass_contrib_multipole}). Recent numerical simulation indeed find indications that support this finding, with the mass bias reaching $b\simeq 0.4$ for massive clusters \citep[e.g.,][]{Henson2017}. However, given the possibly significant contamination from CIB, more investigations and data are needed to conclude in this matter.
The addition of SZ data from ACT and SPT could further help addressing this question, but we leave an exploration to the future.

In this context, we also mention that the expected effect of relativistic SZ \citep{Sazonov1998, Challinor1998, Itoh98, Chluba2012SZpack, Chluba2012moments} is progressively reduced when lowering the $\qcut$. Corrections from relativistic SZ to the distortion shape are not included in the construction of the $y$-map, but should lead to an underestimation of the $yy$-PS, thereby affecting the inferred mass-bias \citep{Remazeilles2019rSZ}. 
For ${\qcut=6}$, the average mass contributing to the SZ PS at $\ell\simeq 100-1000$ drops nearly by a factor of $3$ relative to the total PS calculations (see Fig.~\ref{fig:mean_mass_contrib_multipole}). 
For the standard $yy$-PS analysis, the effect of relativistic SZ on the mass bias is expected to reach the level of $\simeq 1-2$ standard deviations\footnote{The effect is increased when using more recent $T-M$ scaling relations from \citet{Lee2020}.} for \Planck.
Since the bias from relativistic SZ is directly proportional to the power-spectrum weighted mean mass of the sample, we thus expect relativistic SZ to become less important. The proposed completeness modeling of the PS again provides the means to investigate this question in more detail.

\begin{table}
\begin{centering}
\begin{tabular}{lc}
 Mass Bias Measurements

& ($b = $ mean $\pm$ $68\%$ CL)  
\tabularnewline
\hline 
Zubeldia \& Challinor (2019)& $0.29\pm0.10$\tabularnewline
Makyia et al. (2020) & $0.27^{+0.13}_{-0.08}$\tabularnewline
\hline 
Koukoufilippas et al. (2020) & $0.25\pm0.03$\tabularnewline
Chiang et al. (2020)& $0.21\pm0.03$\tabularnewline
\hline 
\Planck Collaboration (2016d) [CNC] & $0.42\pm0.04$\tabularnewline
Bolliet et al. (2018, 2020) [total PS] & $0.40\pm0.05$\tabularnewline
\hline 
\hline 
Total ($q_{\rm cut} \rightarrow \infty$)\; [1-halo] & $0.41\pm0.06$ \tabularnewline
uRC  $q_\mathrm{cut}=20$ \; [1-halo]& $0.34\pm0.06$ \tabularnewline
uRC  $q_\mathrm{cut}=12$\; [1-halo]& $0.32\pm0.06$ \tabularnewline
uRC  $q_\mathrm{cut}=9$\; [1-halo]& $0.29\pm0.05$
\tabularnewline
uRC  $q_\mathrm{cut}=6$\; [1-halo]& $0.25\pm0.05$ \tabularnewline
\hline
uRC  $q_\mathrm{cut}=6$ [1-halo,full MCMC]& $0.13\pm0.05$
\tabularnewline
uRC  $q_\mathrm{cut}=6$ [1+2-halo, full MCMC]& $0.15\pm 0.04$
\tabularnewline
\hline
\hline 
\end{tabular}
\par\end{centering}
\caption{Mass bias measurements, including results from previous works. For detailed discussion see the main text.}
\label{tab:biasconstraints}
\end{table}

\subsection{Signatures of the 2-halo term in the \Planck $y$-map?}
\label{sec:2_halo}
The simple Fisher estimates presented in \sec{sec:total_error_est} suggested the detection of the 2-halo contribution at $\simeq 3.4 \sigma$ after duly accounting for the full trispectrum and marginalization over foregrounds could be possible. Since the 2-halo contribution is only a part of the total model describing \yy, it is not as useful to discuss its stand alone detection. The more relevant quantity is the enhancement in the SNR between the approximate 1-halo modeling and the 1+2-halo modeling of the \yy, for which our Fisher estimates suggest a $\simeq 3.2\sigma$ enhancement (see Table~\ref{tab:snrs}).  Motivated by these observations we now steer our attention to carefully studying the 2-halo contribution to the $yy$-PS and seek its signatures in the \yy estimated from \Planck data for $q_{\rm cut}=6$. We also assess the importance of the 2-halo contribution in the inference of the mass bias.

To detect signatures of this subtle 2-halo contribution, we revise our simplified statistical analysis in favour of more robust approach. The results presented in \sec{sec:mass_bias} relied on using the scaling relation $F_{\rm SZ}$, and assuming this to be valid for different values of $q_{\rm cut}$. We already noted that this scaling relation in detail cannot be expected to hold, as it was derived for the total $yy$-PS and only accounting for the 1-halo prescription \citep{Bolliet2017}. It is therefore not guaranteed to be perfectly valid for $yy$-PS corresponding to different values of $q_{\rm cut}$ and likely will be even more inaccurate when working with the full 1+2-halo model of \yy.

Here we derive the parameter constraints by running the MCMC analysis using the joint CMB-SZ likelihood, which varies all the standard cosmological parameters and mass bias $b$ along with the three SZ foreground template amplitude parameters $(A_{\rm CIB}, A_{\rm RS}$ and $A_{\rm IR})$. Here, while the cosmological parameters of interest $\sigma_8, \Omega_m$ and $H_0$ are primarily constrained by the CMB measurements the bias and SZ foregrounds are constrained by the measurement of the SZ power spectrum. This accounts for all the relevant parameter covariances and data correlations.

We perform two separate analyses, one in which the SZ likelihood uses the 1-halo model and the other in which the complete 1+2-halo model is used to fit the measured SZ spectrum,  simultaneously fitting for the amplitudes of the foreground templates. Both analysis yield cosmological constraints that are consistent with standard cosmology. We first compare the quality of the two fits and find log likelihood improvement of $\Delta {\rm log} \mathcal{L} = -17$ for the 1+2-halo model, indicating a preference towards inclusion of the 2-halo contribution in the theoretical modelling of the SZ-PS. 

\begin{figure}
\centering
\includegraphics[width=\columnwidth]{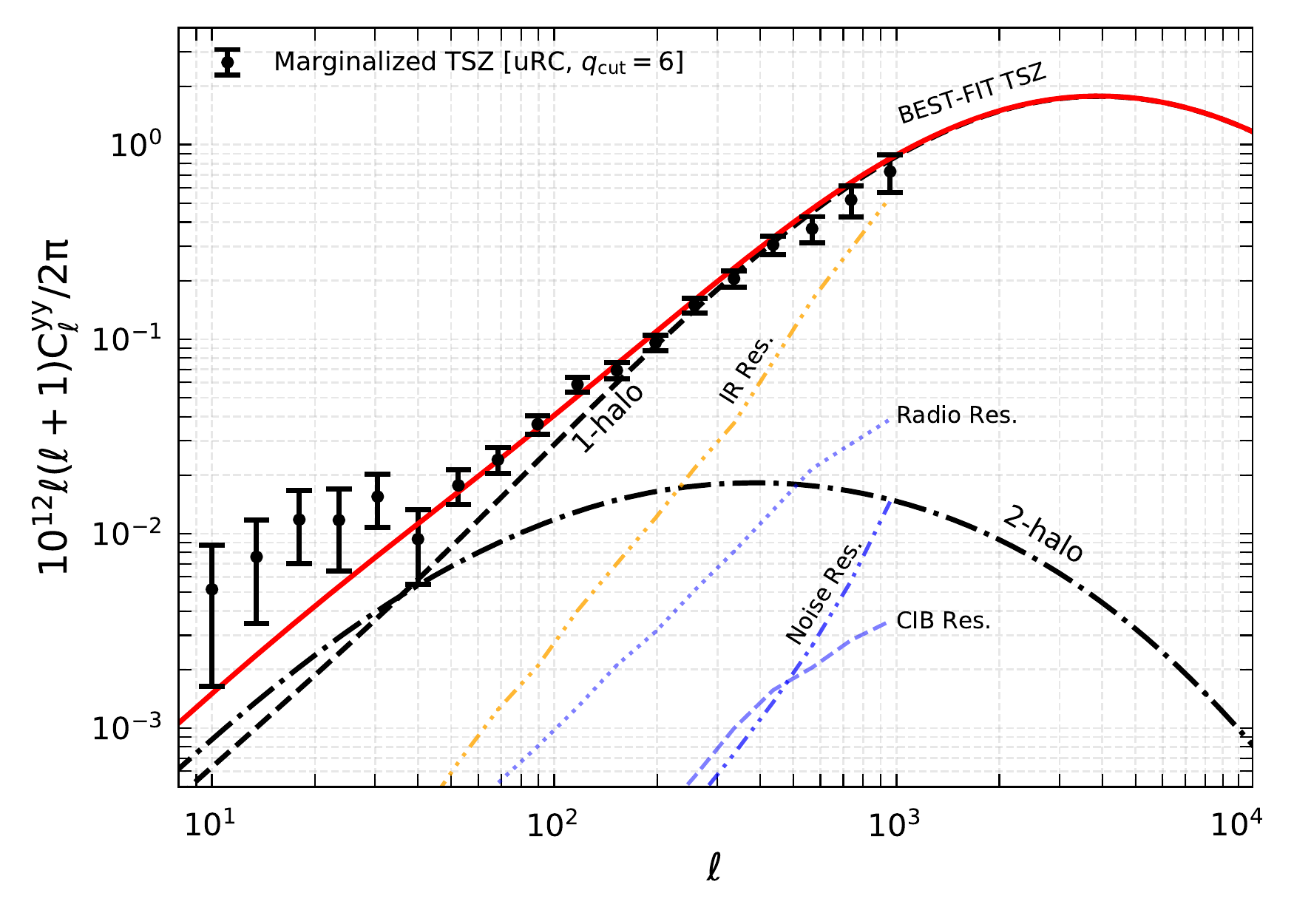}
\vspace{-2mm}
\caption{Marginalized unresolved SZ power spectrum from our analysis with $q_\mathrm{cut}=6$, along with the best-fit model including foreground and noise residuals. Here the best-fitting model is obtained from the analysis of the \Planck 2018 primary CMB data and our $y$-map. The main parameters are $\Omega_{\rm b} h^2=0.022$, $\Omega_{\rm c} h^2=0.12$, $n_\mathrm{s}=0.96$, $\sigma_8=0.82$, $h=0.67$, $B=1.2$ (i.e., $b=0.165$), $A_\mathrm{CIB}=4.71\times 10^{-3}$, $A_\mathrm{IR}=1.49$, $A_\mathrm{RS}=0.19$ and $A_\mathrm{CN}=0.9$. The data for this figure will be made available at \TopoSZ. 
}
\vspace{-2mm}
\label{fig:1h2hfit}
\end{figure}

The best-fit spectrum \yy along with foreground marginalised SZ power spectrum are presented in \fig{fig:1h2hfit}.
We can see that the measured SZ-PS continues to show power in excess of the best fit theory at multipoles below $\ell \lesssim 40$. The SZ contribution from the diffuse warm gas in the local Universe has been suggested to have a nearly constant shape and with an amplitude which closely matches the low-multipole excess seen in the \Planck $yy$-spectrum \citep{Hansen2005}. This contribution is not included in the analytical estimates of the power spectrum and is one possible explanation for the excess power seen at these low multipoles.

Evidently, a more mundane explanation exists:
\Planck $y$-maps are known to suffer from foreground residuals and striping systematics at low multipoles. In addition, in the total $y$-map analysis, the influence of these artefacts is suppressed due to the large trispectrum errors at low multipoles. However, in our topographic analysis, masking high $q$ clusters results in significant reduction in the trispectrum errors, thus making our analysis more prone to large-angle systematics in the $y$-map. 

To quantify the effect of the low-$\ell$ excess further we ran a likelihood analysis in which we exclude the first 34 multipoles (this corresponds to rejecting the first 5 bins in the \Planck binning scheme) of the estimated $yy$-PS. This left our conclusions unchanged, showing that the parameter constraints are mostly driven by scales $\ell\gtrsim 40$. 

\begin{figure}
\centering
\includegraphics[width=0.95\columnwidth]{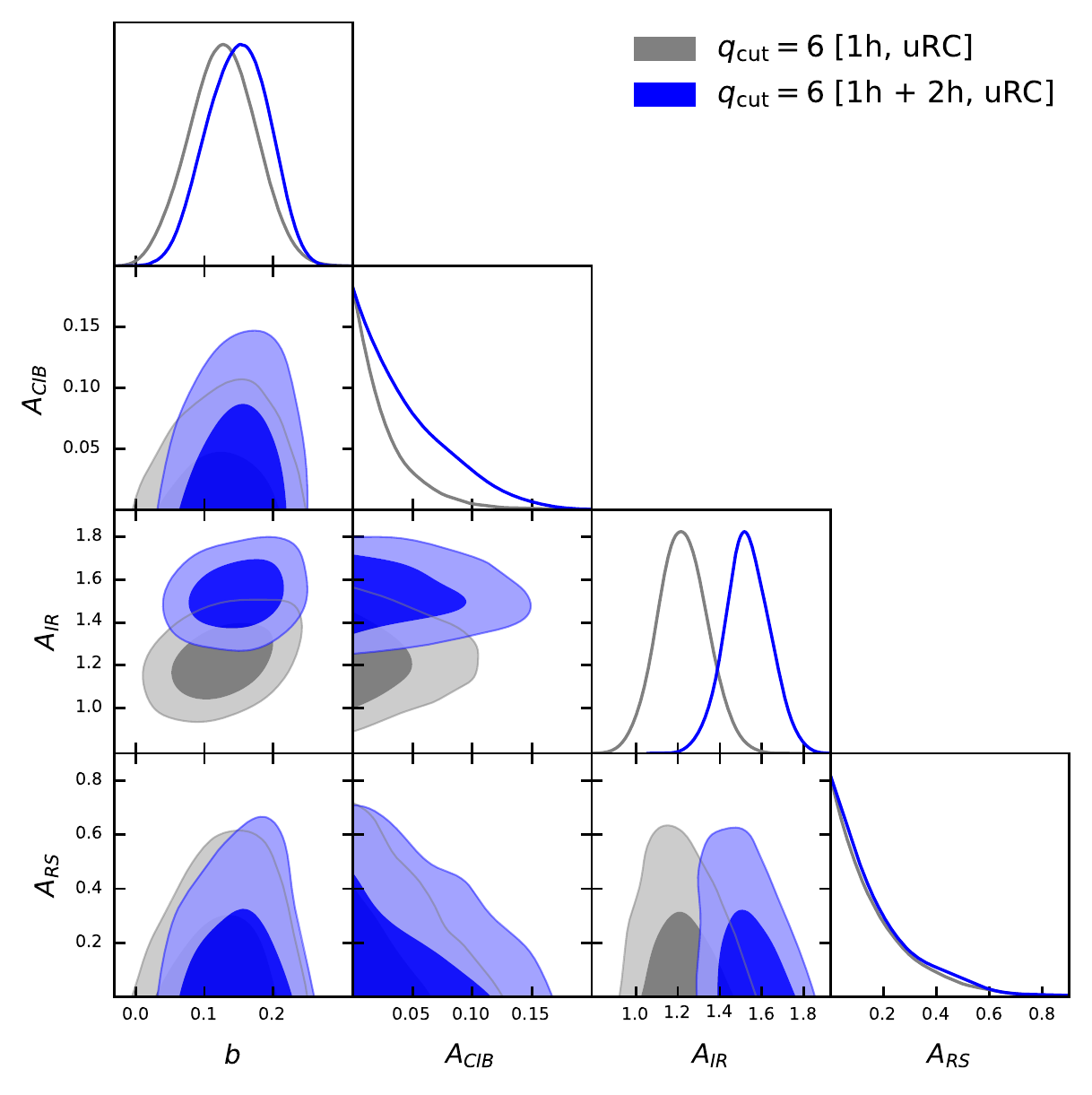}
\caption{Marginalized 2d posterior probability distribution for the analysis with cluster SNR cut $q_\mathrm{cut}=6$.}
\label{fig:1h2hcontours}
\end{figure}

Having noted the importance of the 2-halo component we now return to the discussion of the halo mass bias. The constraints on parameters driven by the SZ measurements for the 1-halo and 1+2 halo likelihood analysis are depicted in \fig{fig:1h2hcontours}. We first note that the constraints on the mass bias when modelling the SZ power spectrum with the 1-halo model are now $b=0.13 \pm 0.05$ as opposed to the value reported with the simplified $F_{\rm SZ}$ treatment, $b=0.25 \pm 0.05$. This confirms that using the effective scaling derived for the total PS becomes inaccurate when working with masked $y$-maps and would have to be re-calibrated for each $\qcut$. 
This difference in analysis also leads to an interpretation that suggest a lower IR contamination in the SZ-PS by nearly a factor of $\simeq 2$.
Finally, we note that when modeling the measured spectrum with complete model yields $b=0.15 \pm 0.04$, which is fully consistent with the value suggested by simulations $b\simeq 0.2$ to within $1.3 \sigma$.

\vspace{0mm}
\section{Discussion and Conclusions}
\label{sec:discussions}
We introduced a new topographic analysis of the Compton $y$-map by decomposing it into a RC (resolved) and uRC (unresolved) component, which is essentially achieved by masking the clusters detected above some preset SNR threshold. We demonstrated that this analysis strategy can be carried out on multi-frequency observations of the microwave sky using only minimal external information. While analogous ideas have been suggested in previous works, these relied of supplementing CMB data with other observations. In addition to introducing the topographic $y$-map analysis, we also use it to analyse the \Planck data, finding several new results.

The topographic analysis suggested here combines the data products resulting from an ILC like analysis that yields the Compton-$y$ map and a multi-matched filter (MMF) analysis that yields characteristics of individual cluster detections (SNR and sky location) as well as the survey completeness function. 
Our formalism then introduces the survey completeness function, traditionally used only in the CNC analysis, into the $y$-map PS modeling (see Sect.~\ref{sec:formalism}). 
Applying this new framework, we demonstrate that masking clusters detected above a given SNR threshold results in a noticeable reduction of the trispectrum errors at a relatively small loss of signal, resulting in an overall enhancement of the SNR of the signal detection. 
Our estimates further demonstrate that the detectability of the 2-halo terms can be enhanced using the suggested strategy (see Fig.~\ref{fig:1halo2halo} and \ref{fig:qcut_err_spec}). Since all these estimates were made accounting for \Planck measurement noise, this suggests that \Planck measurements of the Compton $y$-field could deliver new insights which previous analyses were insensitive to.

Motivated by these observation, we carried out the topographic analysis on the \Planck $y$-map. To make sensible PS estimates for the uRC and RC components we developed an augmented MASTER algorithm (see Sect.~\ref{sec:master_nuances}). Next, we performed a few variants of the likelihood analyses on uRC and RC power spectra derived for different SNR thresholds (see Sect.~\ref{sec:results}). The analysis of the total $y$-map delivered constraints which are consistent with previous findings \citep{Ade:2015fva, Bolliet2017}, serving as a benchmark for our analysis.
We find that progressively reducing the SNR threshold for masking clusters results in a reduction of the $A_{\rm CIB}$ amplitude (see \fig{fig:cosmo_constraint_unresolved}). This is qualitatively consistent with findings of the CIB-tSZ correlation study reported in \citet{planck2016_cib_tsz}. At the same time, we find the mass bias $b$ to systematically decrease, approaching the standard value of $b\simeq 0.2$ found in simulations (see \fig{fig:bias_evol}). This suggests that the tension between cosmological parameters derived from CMB and clusters could be dominantly sourced by the CIB-tSZ correlation.
Alternatively, \Planck data could contain observational hints for mass-dependent bias, as seen in recent hydrodynamical simulation \citep[e.g.][]{Henson2017}. However, it is clear that additional data and analysis are needed to reach a firm conclusion.

In Sect.~\ref{sec:2_halo}, we focus on interpreting the low-multipole excess power observed in the $yy$-PS for all our SNR thresholds. This excess power is also seen in the total $yy$-PS but it becomes more prominent for lower SNR thresholds (see \fig{fig:all_yy_specs}). 
We find indications that adding the 2-halo contribution to the theoretical SZ-PS computation provides a better representation of the data. However, even on including the 2-halo terms, the data still exhibits excess power at the largest angular scales. We speculate that this excess power could be sourced by diffuse SZ \citep{Hansen2005} which is not included in the evaluation of the theoretically $yy$-power spectrum; however residual foreground contamination is likely to play a crucial role, warranting a more careful reanalysis starting from the raw data products of \Planck.
By excluding the first 34 multipoles of the SZ-PS from the analysis, we verified that our conclusions on the mass bias are not affected by this excess.

Our novel framework for the $yy$-PS analysis comes with several important benefits. Since the same completeness modeling is an essential ingredient to the corresponding CNC analysis, we can now confidently combine this with the SZ-PS likelihood.
The natural idea is to use the PS modeling for the uRC component while treating the RC component using CNC. Since the 2-halo term is a correction to the RC component, the two data products are largely independent. By repeating the analysis with varying $\qcut$ we can study the dependence of derived parameters on the chosen split, introducing a powerful new diagnostic. This can in principle be used to obtain a mass-dependent\footnote{In detail a precise weighting scheme has to be introduced to compute the effective mass relevant to the observable.} measurement of the bias parameter.
In the future, with the advent of more data from the SO \citep{SOWP2018} or CMB Stage-IV \citep{Abazajian2016S4SB}, this can be further refined, potentially even allowing to introduce $\qcut$-bins. 
Given the important role of CIB contamination, high frequency coverage provided by CCAT-prime \citep{CCATp2018} could become highly relevant. 
The method could be additionally enhanced by adding information from higher order statistics, which can help break degeneracies, as has recently been demonstrated  \citep{Ravenni2020}.
Together this would provide a powerful way to model the underlying bias parameter and halo-mass function.

We finally emphasize that the uRC component contains new complementary information with respect to the RC. Indeed, for low $\qcut$, it is largely impossible to speak of separate clusters. This means that the $yy$-PS can become sensitive to new contributions such as diffuse SZ \citep{Hansen2005}. 
Since this contribution is expected to dominate at large angular scales, it will be very interesting to further study this signal using space-based experiments such as {\it Litebird} \citep{Litebird}, {\it PIXIE} \citep{Kogut2011PIXIE, Kogut2016SPIE} or more futuristic CMB imagers \citep{Hanany2019PICO, Basu2019Voyage}. Since accessing the low multipoles from the ground is challenging, this could be an exciting target for space-based experiments.

\section*{Acknowledgments}
This work was supported by the ERC Consolidator Grant {\it CMBSPEC} (No.~725456) as part of the European Union's Horizon 2020 research and innovation program.
Some results in this work are derived using  \href{https://healpy.readthedocs.io/en/latest/#}{ \rm healpy} \citep{healpy} and \href{https://healpix.jpl.nasa.gov}{ \rm Healpix} \citep{healpix}. 
JC was furthermore supported by the Royal Society as a Royal Society University Research Fellow at the University of Manchester. We would like to thank Colin Hill and Richard Battye for many insightful exchanges as well as Andrea Ravenni, Eiichiro Komatsu, Ryu Makyia, David Alonso, Etienne Pointecouteau and Monique Arnaud for insightful discussions and cross-checks related to pressure profile computations. We are also grateful to Thejs Brinckmann for help with \verb|MontePython|, Anthony Lewis, Jesus Torrado and Tim Morton for help with \verb|cobaya|. We are also very grateful to Anthony Holloway and Sotirios Sanidas for continuous technical support with our computing hardware at JBCA.

\section*{Data Availability} The data products of this paper will be made available at \TopoSZ under \url{https://github.com/CMBSPEC/TopoSZ.git}. All Planck data used in this work are publicly available on the \href{http://pla.esac.esa.int/pla}{Planck Legacy Archive}.

\bibliographystyle{mnras}
\bibliography{myref,ref,Lit} 

\begin{appendix}
%
\section{The 1pPDF of the masked $y$-maps}
\label{app:1ppdf_study}
Ideally the Compton $y$-map is expected to be fully positive, since it is a measure of the electron gas thermal pressure integrated along the line of sight. However the $y$-map inferred from analysis on multi-frequency microwave maps does not have this property, owing to contamination from foregrounds and measurement noise \citep[e.g.,][]{Jose2003PDF}. The negative excursions in the 1-point PDF (1pPDF) due to measurement noise can be estimated from differences between $y$-maps reconstructed using the half mission 1 and half mission 2 data sets. Any excess negative excursions in the 1pPDF of the reconstructed $y$-map, over and above those expected from noise alone, must be due to residual foregrounds.
\begin{figure}
\centering
\includegraphics[width=0.95\columnwidth]{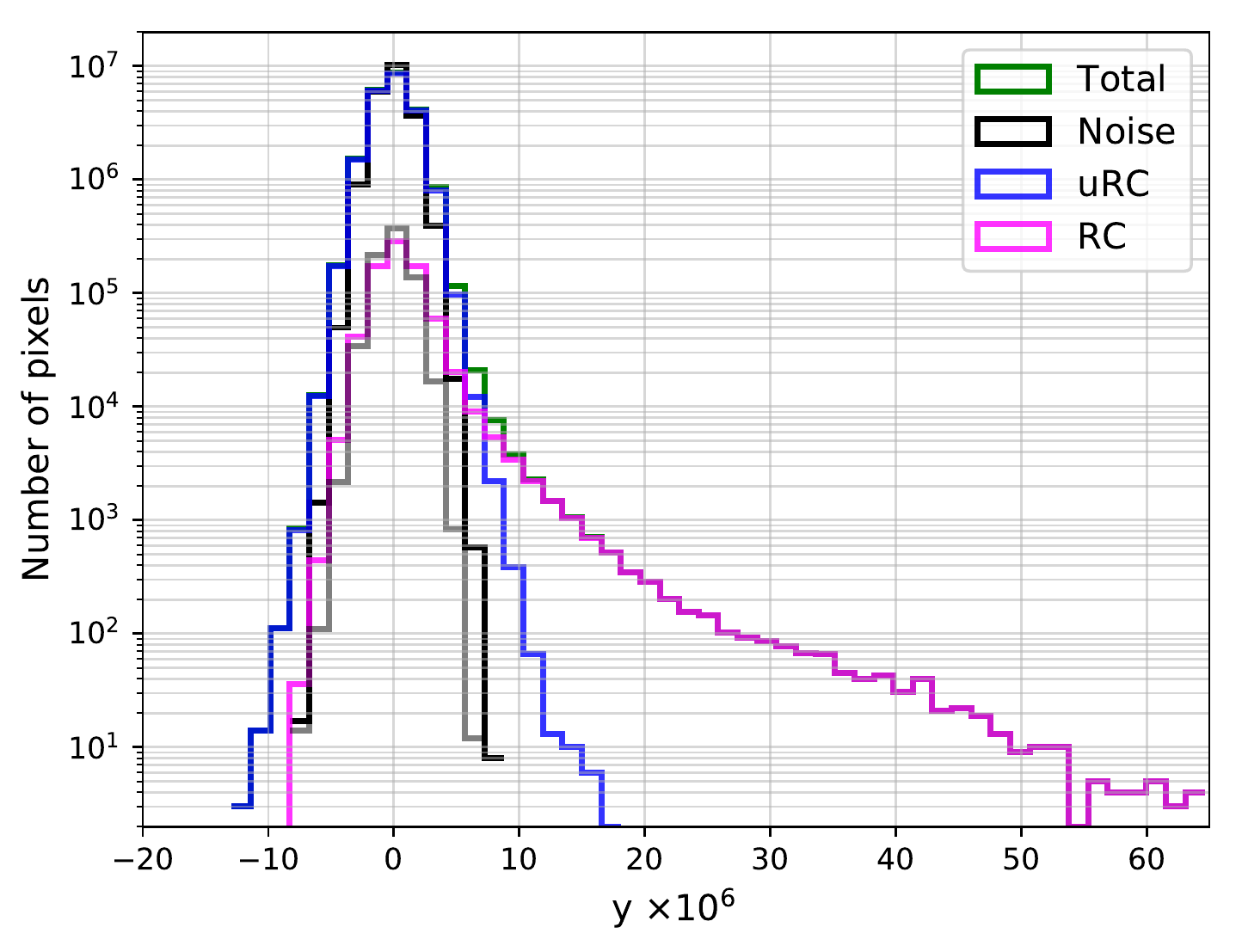}
\caption{This figure depicts the un-normalized histogram of total, uRC and RC component of the NILC $y$-map for $\qcut=6$. The black histogram is that of the measurement noise while the green histogram is that of the NILC $y$-map  in the analysis region defined by the mask $\mathcal{M}_G$. The blue histogram depicts the histogram of the uRC component of the $y$-map and notably it overlays the negative excursion seen in the green histogram. The gray histogram depicts the un-normalized noise histogram while the magenta histogram is that of the RC component defined by $\mathcal{M}_{\rm RC}$. Notably the magenta histogram overlays the positive end of the green histogram and appears consistent with noise on the negative end.}
\label{fig:nilc_histogram}
\end{figure}

Figure~\ref{fig:nilc_histogram} depicts the histogram of the of the reconstructed $y$-map and the corresponding noise for different analysis masks. The black (gray) curves depicts the 1pPDF of the noise in regions corresponding to the uRC (RC) component of the reconstructed $y$-map. Subtracting the RC component from the $y$-maps, results in the blue histogram which has a weaker positive skewness tail, since the high mass cluster contribution to the $y$-map have been removed, Gaussianizing the distribution of the reconstructed $y$-map. However note that the negative excursion is identical to the excursion seen in the green histogram corresponding to the total $y$-map, indicating that a dominant portion of the galactic foreground contamination is in the uRC of the $y$-map. The remnant excess positive skewness seen in the blue histogram can be attributed to un-subtracted clusters and some foreground residuals in the $y$-map. 

The 1pPDF corresponding to the RC component of the $y$-map has a negative tail, which appears consistent with the excursion expected from measurement noise, while almost completely accounting for the positive skewness originally seen in the green histogram (capturing a dominant fraction of the non-Gaussian peaks in the $y$-field). While galactic and extra-galactic foregrounds (e.g. CIB) can also add in positive to the $y$-map, these cannot be simply diagnosed by inspecting the 1pPDF of the $y$-map and requires a more detailed model-dependent analysis which is discussed in the main text.

\section{Master algorithm to measuring the PS in the presence of a mask}
\label{app:test_master}
%
\begin{figure}
\centering
\subfigure[Un-resolved cluster mask]{\includegraphics[width=1\columnwidth]{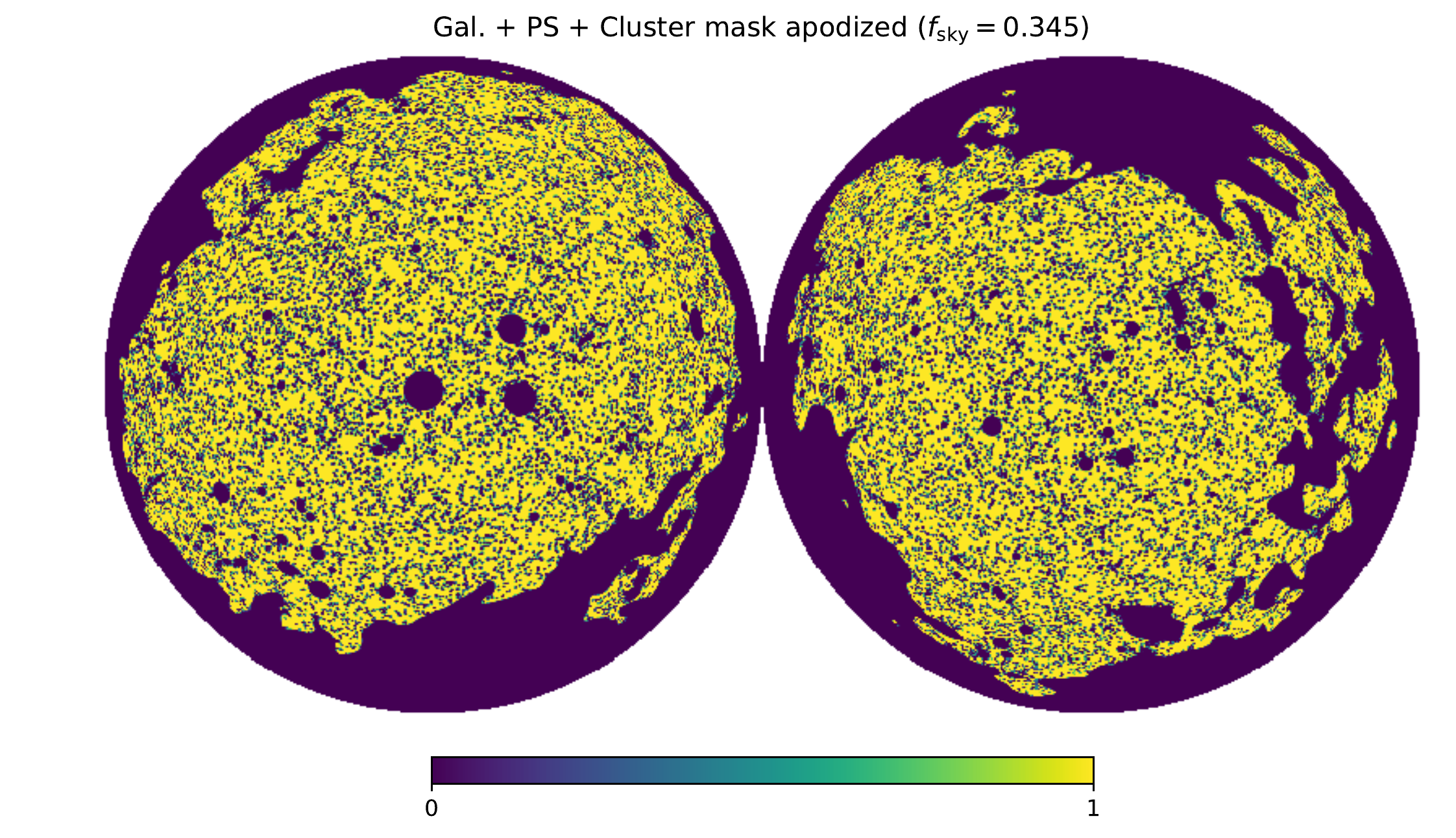}}
\subfigure[Resolved cluster mask]{\includegraphics[width=1\columnwidth]{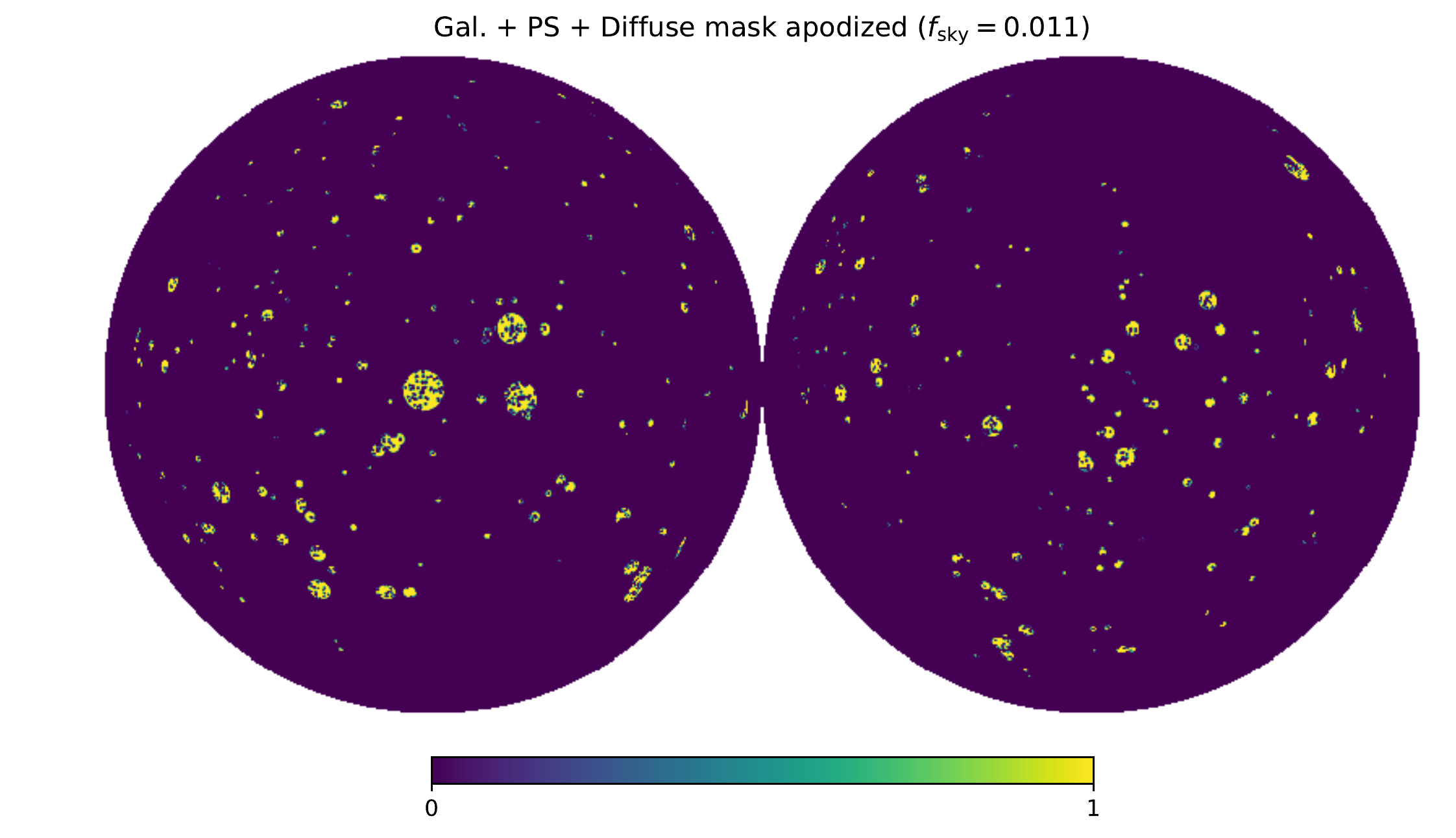}}
\caption{The  mask depicted on top is used to estimate the contribution to the power spectrum sourced by the uRC while that in the bottom is  used to estimate the contribution from the RC of the Compon y-map.}
\label{fig:masks}
\end{figure}
\begin{figure}
\centering
\includegraphics[width=1\columnwidth]{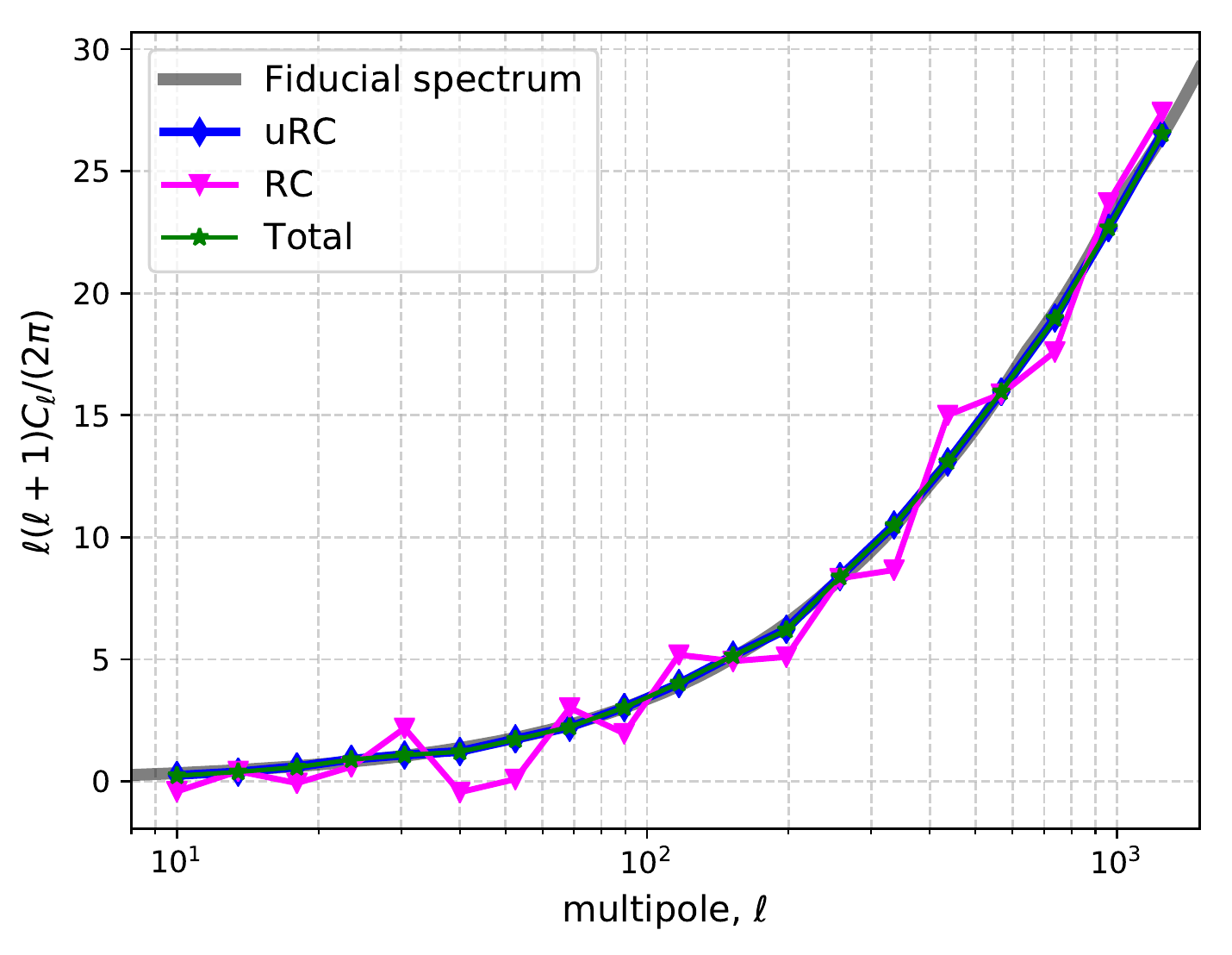}
\caption{The corrected power spectrum estimated using the conventional MASTER algorithm for masks used total, RC and uRC analysis. Note that the recovered spectra in the three different analysis yield power spectra which are consistent with the injected fiducial spectrum.}
\label{fig:test_master}
\end{figure}

The masks used in the analysis are fairly complicated. This is particularly true while estimating the PS contribution from the RC component. The respective masks are depicted in \fig{fig:masks}.
To ensure our implementation of the MASTER algorithm works as expected even on this aggressive RC mask we ran several null tests. We simulate a Gaussian realization of the $y$-map using the fiducial Compton $y$-map power spectrum. To this simulation we apply the masks corresponding to the uRC and RC components of the $y$-map. Finally we estimate the master corrected power spectrum of the respective fields and the results from this exercise are summarized in \fig{fig:test_master}. This exercise confirms that our implementation of the MASTER algorithm is able to make an unbiased recovery of the input fiducial power spectrum even for the very aggressive mask used to estimate the contribution from the resolved clusters.
%

\section{Treatment of the parameter dependence of the covariance matrix}
\label{app:test_cov}

\begin{figure}
\centering
\includegraphics[width=1\columnwidth]{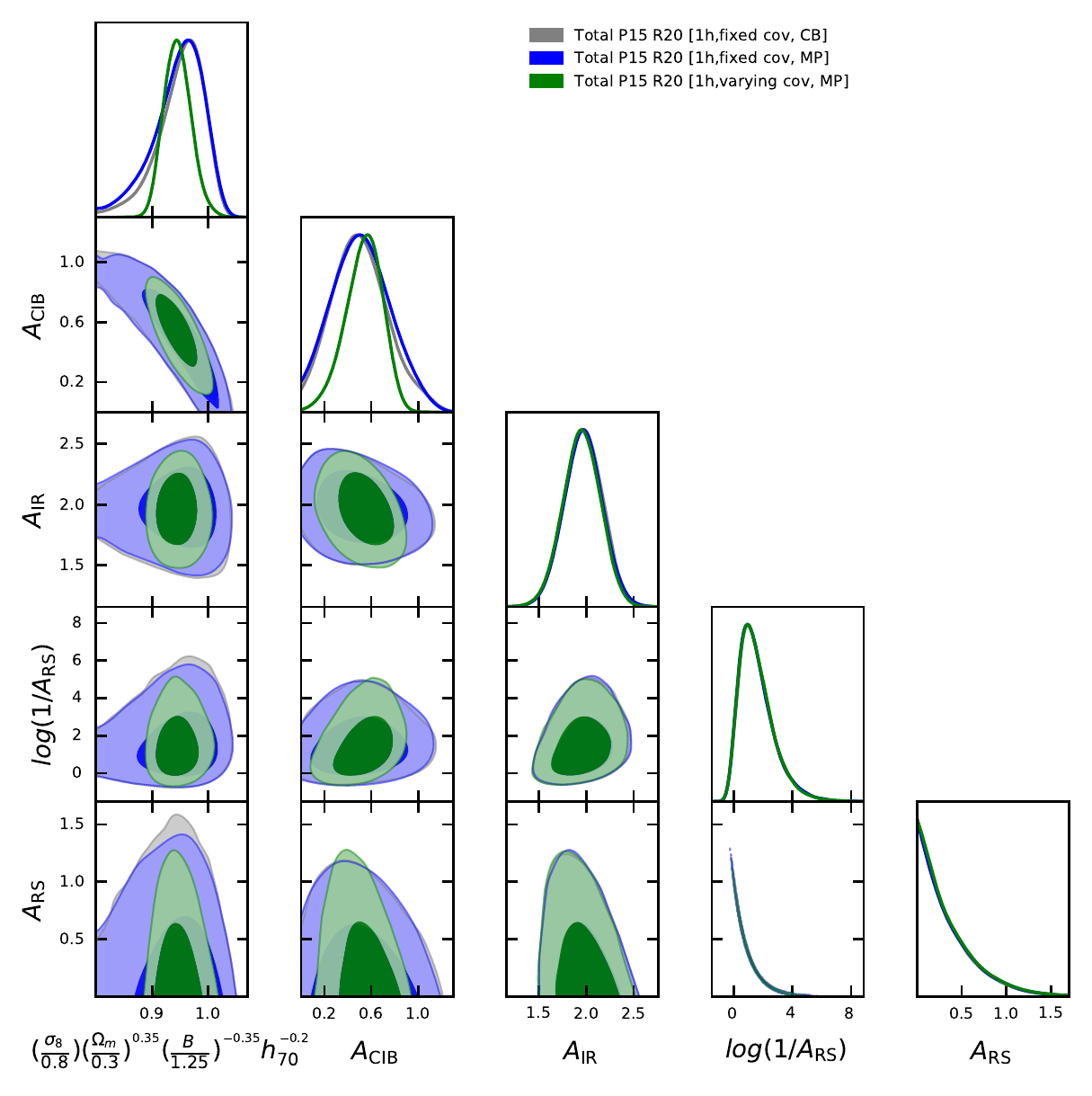}
\caption{Comparison of the posteriors for different methods and codes to perform the maximum likelihood analysis for the total $yy$-PS case. {\tt MontePython} [MP] and {\tt Cobaya} [CB] agree well. The fixed covariance matrix approach followed here increases the width of the posteriors when compared to the varying covariance matrix approach.} 
\label{fig:compmcmc_full}
\end{figure}

\begin{figure}
\centering
\includegraphics[width=1\columnwidth]{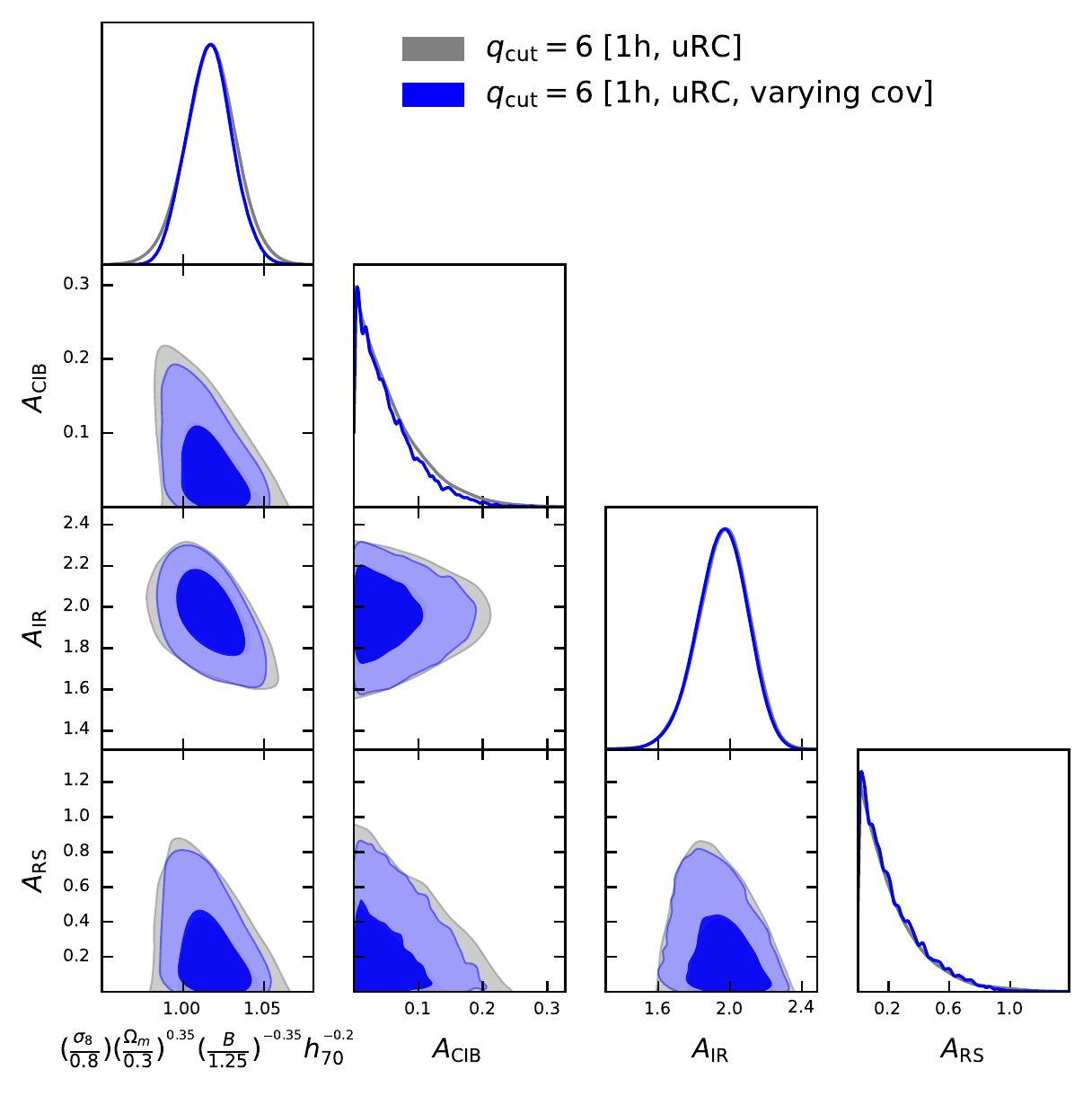}
\caption{Comparison of the posteriors for $\qcut=6$ with and without varying covariance matrix. This choice has a marginal effect on the result.} 
\label{fig:compmcmc_6}
\end{figure}

We explored the possibility of fixing the covariance matrix and combining with an iterative approach, as suggested in \cite{Makiya:2019lvm}. This has been argued to change the results of the inference for Gaussian fields \citep{Carron_2013}. 
Following \cite{Makiya:2019lvm}, here we keep the non-Gaussian covariance fixed in the likelihood evaluation. The latter, is estimated from some fiducial set of parameters while only updating the theoretical $C_{\ell}^{yy}$ power spectrum. The parameter optimization procedure is repeated multiple times, using the best-fit parameters inferred from the previous iteration to define the covariance matrix and this is repeated until convergence. 

For the total PS analysis, this increased the final error on $b$ by a factor of $\simeq 4$, i.e., $b=0.46\pm 0.21$, by introducing heavy wings to the posteriors of $F_{\rm SZ}$ (see Fig.~\ref{fig:compmcmc_full}). Once completeness modeling is included, we find a small effect when using fixed versus varying covariance (see Fig.~\ref{fig:compmcmc_6} for $\qcut=6$). Our main results are thus not affected by these differences.

\begin{table}
\centering
\begin{tabular}{ccc}
\toprule
Parameter &   min. &  max. \\
\midrule
             $h$ &     0.55 &     0.90 \\
             $\Omega_{\rm b} h^2$ &     0.020 &     0.025 \\
             $\Omega_{\rm c} h^2$ &     0.11 &     0.13 \\
           $10^{9} A_\mathrm{s}$ &     0.1 &     10 \\
        $n_\mathrm{s}$ &     0.94 &     1 \\
            $B$ &     1 &     2 \\
        $A_\mathrm{CIB}$ &     0 &     5 \\
        $A_\mathrm{IR}$ &     0 &     5 \\
        $A_\mathrm{RS}$ &     0 &     5 \\
\bottomrule
\end{tabular}
\caption{Range of the uniform prior probability distribution used in our maximum likelihood analyses.}
\label{tab:priors}
\end{table}

\end{appendix}

\bsp	
\label{lastpage}
\end{document}